\documentclass[aps,preprint]{revtex4-1}

\usepackage{csquotes}
\usepackage{float}
\usepackage{pgfplotstable}
\usepackage{amsmath,amsthm,verbatim,amssymb,amsfonts,amscd,relsize}
\usepackage{graphicx}
\usepackage{gensymb}
\usepackage{csvsimple,booktabs}
\usepackage[export]{adjustbox}
\usepackage{capt-of}
\usepackage{url}
\begin{document}

\title{Thermodynamics of mixtures of patchy and spherical colloids of different sizes: a multi-body association theory with complete reference fluid information}
\author{Artee Bansal}\email{ab43@rice.edu}
\author{Arjun Valiya Parambathu}\email{av42@rice.edu}
\author{D.~Asthagiri}\email{dna6@rice.edu}
\author{Kenneth R. Cox}\email{krcox@rice.edu}
\author{Walter G. Chapman}\thanks{wgchap@rice.edu}
\affiliation{Department of Chemical and Biomolecular Engineering, Rice University, Houston}
\date{\today}
\vfill
\pgfkeys{/pgf/number format/.cd,precision=5}
\begin{abstract}
We present a theory to predict the structure and thermodynamics of mixtures of colloids of different diameters, building on our earlier
work [J. Chem. Phys. 145, 074904 (2016)] that considered mixtures with all particles constrained to have the same size. The patchy, solvent 
particles have short-range directional interactions, while the solute particles have short-range isotropic interactions. The hard-sphere
mixture without any association site forms the reference fluid. An important ingredient within the multi-body association theory is
the description of clustering of the reference solvent around the reference solute. Here we account for the physical, multi-body clusters of 
the reference solvent around the reference solute in terms of occupancy statistics in a defined observation volume.  These
occupancy probabilities are obtained from enhanced sampling simulations, but we also present statistical mechanical models to estimate these probabilities with limited simulation data. Relative to an approach that describes only up to three-body correlations in the reference, incorporating the complete reference information better predicts the bonding state and thermodynamics of the physical solute for a wide range of system conditions.   Importantly, analysis of the residual chemical potential of the infinitely dilute solute from molecular simulation and theory shows that whereas the chemical potential is somewhat insensitive to the description of the structure of the reference fluid the energetic and entropic contributions are not, with the results from the complete reference approach being in better agreement with particle simulations. 
\end{abstract}

\maketitle

\section{Introduction}

 In thermodynamic perturbation theory of association involving short range interactions between molecules, the properties of the reference
 fluid plays a central role. In the typical situation when the reference is a hard-sphere fluid,  perturbation theories usually 
 use information about two body, and at times three body, correlations in the reference fluid to describe the physical (associating) system. For example, Wertheim's theory \cite{wertheim_fluids_1984,wertheim_fluids_1984-1} and its extensions based on the statistical associating fluid theory (SAFT) \cite{chapman_new_1990} use pair correlation information at contact to estimate extent of association between pairs of molecules. 
In SAFT, for the hard-sphere reference either the Carnahan-Starling \cite{carnahan_equation_1969} equation for a single component fluid 
or the Boublik-Mansoori-Carnahan-Starling-Leland \cite{boublik_hardsphere_1970,mansoori_equilibrium_1971} equation for a mixture
are used to describe the pair-correlation information at contact. The structure of Wertheim's theory or SAFT is such that one can obtain
accurate extent of association and thermodynamics even for systems with strong inter-particle interactions provided 
the representation of the reference is adequate.  However, as the complexity of the interaction increases in the physical system, such as may result from 
multiple bonding and size asymmetries,  information about two or three body correlations in the reference no longer suffices. 
 
In our previous work \cite{bansal_structure_2016}, we studied the multi-body correlation functions of a symmetric hard sphere reference fluid in terms of the probabilities of observing $n$ molecules in the bonding region.  These occupancy probabilities were obtained from enhanced sampling Monte Carlo simulations for the hard sphere fluid. We developed a procedure to use this information within the Marshall-Chapman formalism \cite{marshall_molecular_2013,marshall_thermodynamic_2013} to describe multiple association of solvent molecules to a solute molecule.   This
\textit{complete reference} approach proved successful in predicting the bonding state and thermodynamics of a colloidal solute in a patchy solvent for a wide range of system conditions \cite{bansal_structure_2016}. 

Here we study mixtures where the solute diameter is as small as half to as large as twice the diameter of the solvent.  The solvent particles are spheres with directional interaction sites and the solute particles are spheres with isotropic interactions, and  the solute is capable of bonding with multiple solvent particles. The structure and thermodynamics of mixture of hard spheres with different diameters
has been studied in detail before \cite{torquato_microstructure_1986,reiss_statistical_1959,mayer_integral_1947,torquato_microstructure_1982,torquato_microstructure_1983,torquato_microstructure_1985}, but a compact form for the correlations beyond the contact value is still unavailable. Further, for systems with large size asymmetries even the pair-correlation information obtained using the Boublik-Mansoori-Carnahan-Starling-Leland equation is inadequate \cite{feng_contact_2011}.  Our approach of including multi-body correlations rests on using the occupancy statistics \cite{reiss_upper_1981,pratt_quasichemical_2001,pratt_selfconsistent_2003,bansal_structure_2016} of the hard-sphere solvent around the 
hard-sphere solute. We find that representing multi-body correlation functions in terms of occupancy statistics in physically reasonable observation volumes accurately captures the multi-bonding effects in asymmetric mixtures.  These occupancy statistics are  obtained from particle simulations. Importantly,  we also present a physically transparent, statistical mechanical model to describe the occupancy probabilities in symmetric and asymmetric hard sphere fluids for different packing fractions. This model corrects multi-body effects obtained for isolated clusters by incorporating the role of the cluster-bulk interface and the bulk medium effects. We also investigate the energy-entropy decomposition of the chemical potential of the solute in a model system with only solute-solvent interactions to better appreciate the role of the reference fluid. Throughout, theoretical results are validated versus molecular simulations

The rest of the paper is organized in the following way.  In Section~\ref{sc:bentheory} we discuss the association potential of the system and describe how packing effects are important for the given potential.  The Marshall-Chapman \cite{marshall_thermodynamic_2013} theory is briefly introduced to show the multi-density representation of the free energy, and based on our previous work \cite{bansal_structure_2016}, an improved representation of multi-body correlations(\textit{complete reference}) is presented. In Section~\ref{sc:HStheory}, we examine hard sphere packing around a reference particle and develop models based on statistical mechanics \cite{reiss_upper_1981} and hard sphere simulation data for different densities. We apply our complete reference approach for different asymmetric mixtures of solute and solvent and present results in Section~\ref{sc:res_asso}.
In Section~\ref{sc:res_hs} we present results for the hard sphere reference system (symmetric and asymmetric mixtures) based on the correlation developed in section~\ref{sc:HStheory}. We also provide simulation results for  isolated cluster probabilities in asymmetric hard sphere mixtures in the appendix (Section~\ref{sc:appen}). 

\section{Theory} 
\subsection{Asymmetric mixtures with different association geometries }  \label{sc:bentheory}
The focus of our study is asymmetric mixtures containing molecules with short range attractive interactions. The short range association potential is the same as that in previous work \cite{bansal_structure_2016}: the solute molecule  can associate with multiple solvent molecules isotropically and the patchy solvent has directional interactions. 
The total potential is a sum of hard sphere and association contributions 
\begin{equation}
u{(r)}=u_{HS}{(r)}+u_{AS}{(r)}
\label{eq:potT}
\end{equation}

\begin{figure}[h!]
	\includegraphics[scale=0.5]{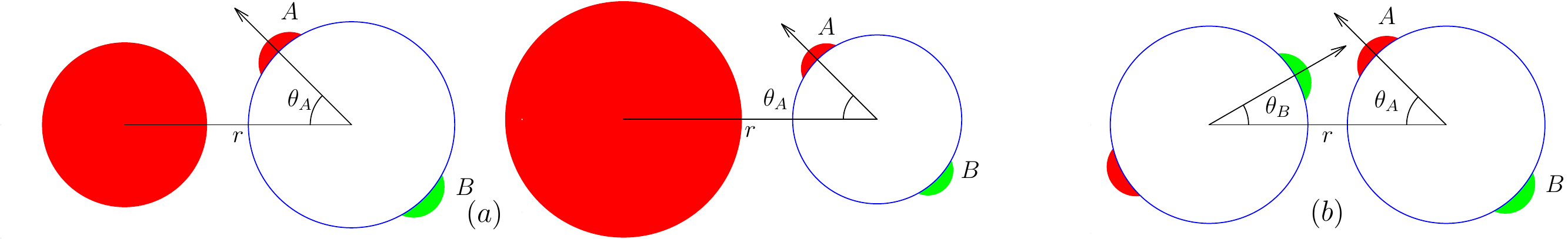}
	\caption{Association between solute and solvent (a) and solvent molecules (b). Different Cases with solute larger (middle) and smaller (left) than solvent molecules are studied. $r$ is the center-to-center distance and $\theta_A$ and $\theta_B$ are the orientation of the attractive patches $A$ and $B$ relative to line connecting the centers. The solute (colored red) can only interact with patch $A$ (colored red).}\label{fig:1}    
\end{figure}

The association potential for patchy-patchy $(p,p)$  and spherical-patchy $(s,p)$ particles is:
  \begin{equation}
  	u_{AB}^{(p,p)}{(r)}=
  	\begin{cases}
  		-\epsilon_{AB}^{(p,p)}, r<r_c \,\text{and}\, \theta_A\leq \theta_c^{(A)}\,\text{and}\,\theta_B \leq \theta_c^{(B)}
  		\\
  		0   \text{ \ \ \ \ \ otherwise}
  		\\    
  	\end{cases}
  	\label{eq:potential1}
  \end{equation}
  
  \begin{equation}
  	u_A^{(s,p)}{(r)}=
  	\begin{cases}
  		-\epsilon_A^{(s,p)}, r<r_c\, \text{and}\, \theta_A\leq \theta_c^{(A)}
  		\\
  		0   \text{ \ \ \ \ \ otherwise}
  		\\    
  	\end{cases}
  	\label{eq:potential2}
  \end{equation}
   where the subscripts $A$ and $B$ represent the type of site and $\epsilon$  is the association energy; $r$ is the distance between the particles; and $\theta_A$ is the angle between the vector connecting the centers of two molecules and the vector connecting association site $A$ to the center of that molecule (Fig.~\ref{fig:1}).  The critical distance beyond which particles do not interact is $r_c$ and $\theta_c$ is the solid angle beyond which sites cannot bond. Fig.~\ref{fig:1} shows examples of solute-solvent and solvent-solvent short range interaction geometries for different sizes of solute particles.   

\begin{figure*}[h!]
	\centering
	\includegraphics[scale=0.5]{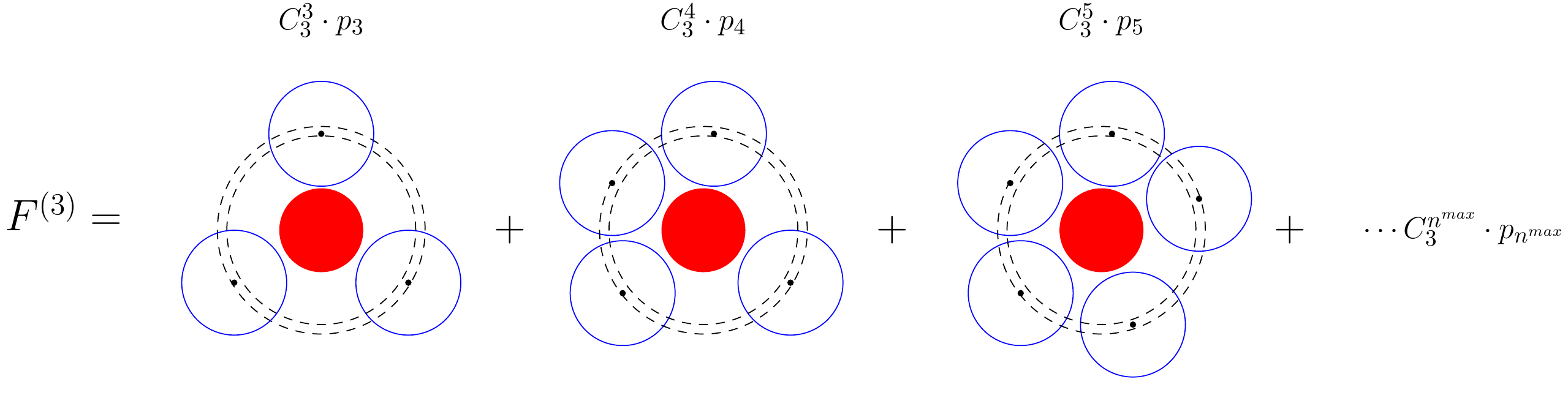}
	\caption{Schematic of $F^{(n)}$, the average number of $n$-solvent cluster ($n$-mer) around a solute (Eq.~\ref{eq:Fn}), for $n=3$. $F^{(n)}$ for the reference fluid is an important target of our study. $p_m$ is the probability of observing exactly $m$-solvents in the observation volume around the solute. $C^m_n$ is the binomial coefficient. The region between the two concentric dashed circles of radii, $\bar \sigma$ (average diameter) and $r_c > \bar \sigma$, respectively, represents the bonding volume. }
	\label{fig:Fn} 
\end{figure*}

Since the solute can associate with multiple solvent molecules (Eq.~\ref{eq:potential2}), it is important to study the multi-body correlations that determine the packing of solvent particles around the solute in the reference fluid  \cite{bansal_structure_2016}. The difficulty in determining these interactions arises due to the limited knowledge in describing multi-body correlation functions for $n\ge 3$. 
But the volume integral of the multi-body correlation has a clear physical meaning in terms of average number of $n$-solvent clusters ($F^{(n)}$, Fig.~\ref{fig:Fn}). In particular, for the distinguished solute, 
\begin{eqnarray}
{F^{(n)}} & = &\frac{{{\rho_p^n}}}{{n!}}\int\limits_{v} {d{{\vec r}_1} \cdots \int\limits_{v} d{{\vec r}_n}{g_{HS}}\left( {{{\vec r}_1} \cdots {{\vec r}_n}|0} \right)} \nonumber \\
& = & {\sum\limits_{m = n}^{{n^{\max }}} {C^m_n p_m}} \, ,
\label{eq:Fn}
\end{eqnarray}
where $\rho_p$ is the density of solvent particles, $p_n$ is the probability of observing exactly $n$ solvent particles
in the observation volume of the solute ($v$) defined by the spherical region of radius $r_c$, $C^m_n$ ($=m! / (m-n)!\cdot n!$), and 
$g_{HS}({\vec r}_1 \cdots {\vec r}_n|0)$ is the distribution function of the $n$-solvent particles around the solute at the center of the observation volume, indicated by $(\ldots | 0)$. $n^{max}$ is the maximum number of solvent molecules that can occupy the observation volume around the reference solute.
  

In Wertheim's multi-density formalism \cite{wertheim_fluids_1986,wertheim_fluids_1986-1}, the free energy due to association ($A^{AS}$) is
expressed as
  \begin{equation}
  \frac{{A^{AS}}}{{V{k_{\rm B}}T}} = \sum {\left( {{\rho_k}\ln \frac{{\rho ^{(0)}_{k }}}{{{\rho_{ k }}}} + {Q^{\left( k \right)}} + {\rho _{ k }}} \right)} - 	\frac{	\Delta c^{(0)}}{V}
  \label{eq:3}
  \end{equation}
  
  where $k_{\rm B}$ is the
  Boltzmann constant, $T$ is the temperature, the summation is over the species ($k={s,p}$),  $\rho$ is the number density, $\rho^{(0)}$ is the monomer density, $Q^{(k)}$ is obtained from the Marshall-Chapman development \cite{marshall_thermodynamic_2013} and  $\Delta{c^{0}}$ is the contribution to the graph sum due to association between the solvent-solvent  $(p,p)$  and solute-solvent $(s,p)$ molecules, i.e.
  \begin{equation}
  \Delta c_{}^{\left( 0 \right)} = \Delta c_{pp}^{\left( 0 \right)} + \Delta c_{sp}^{\left( 0 \right)}
  \label{eq:4}
  \end{equation}
 
Marshall and Chapman \cite{marshall_molecular_2013,marshall_thermodynamic_2013}  extended Wertheim's theory beyond the single bonding condition to incorporate multi-body effects in a solution consisting of an isotropic solute and solvent with directional interactions. The contribution to free energy due to association between solute and solvent molecules was obtained as 
 \begin{equation}
  \frac	{\Delta c_{sp}^{(0)}}{V} = \sum\limits_{n = 1}^{{n^{\max }}} {\frac{\Delta c_n^{( 0)}}{V} }
 	\label{eq:5}
 \end{equation}
 where the sum is over different coordination states of the solute and $\Delta c_n^{(0)}$ is given by:
 \begin{widetext}         
 	\begin{eqnarray}
 		\Delta c_n^{(0)} & = & \frac{\rho ^{(0)}_s {( {\rho_p X_A^{(p)}})}^n} {\Omega^{n + 1} n!} \int d(1)\cdots d(n + 1) \,
 		g_{HS}( 1 \cdots n + 1) \cdot \prod\limits_{k = 2}^{n + 1} {( f_{A}^{(s,p)} ( 1,k))}  \, .
 		\label{eq:14}
 	\end{eqnarray}
 \end{widetext}

 In Eq.~\ref{eq:14}, $\rho_p=\rho\cdot x^{(p)}$ is the density of solvent molecules obtained from the mole fraction of solvent($x^{(p)}$) and the total density($\rho$), $X_A^{(p)}$ is the fraction of solvent molecules not bonded at site A,  $\Omega =4\pi$ is the total number of orientations, $f_{A}^{( {s,p})}(1,k) = (\exp (\varepsilon_A^{(s,p)}/k_BT) - 1)$  is the  Mayer function for association between $p$ and $s$ molecules corresponding to potential in Eq.~\ref{eq:potential2} and the integral is over all the orientations and positions of the $n+1$ particles. By taking the average association strength and acceptable orientations out of the integral and fixing the solute at the origin, the above integral can be rewritten as  
 \begin{widetext}
 	\begin{eqnarray}
 		\frac{	\Delta c_n^{(0)}}{V} & = & \frac{\rho _s^{(0)} {( {{\rho}_{p}X_A^{(p)}}f_A^{(s,p)}\sqrt {\kappa _{AA}})}^n} {n!} \int_{v} d\vec r_1 \cdots  \int_{v} d\vec r_n \,
 		g_{HS}(\vec r_1 \cdots \vec r_n |0)  \, .
 		\label{eq:3}
 	\end{eqnarray}
 \end{widetext}
Marshall and Chapman\cite{marshall_molecular_2013,marshall_thermodynamic_2013} approximated the integral in Eq.~\ref{eq:3} as 
  \begin{eqnarray}
  \int_{v} d\vec r_1 \cdots \int_{v} d\vec r_n\, g_{HS}(\vec r_1 \cdots \vec r_n |0)  \approx y_{HS}^n( \sigma) \delta ^{(n)} \Xi ^{(n)} \, ,
  \label{eq:MCA}
  \end{eqnarray}
where  $\Xi^{(n)}$ is the partition function for an isolated cluster of $n$ solvent hard-spheres around a solute hard-sphere, 
$y_{HS}( \sigma)$ is (pair) cavity correlation function at contact, and $\delta^{\left(n\right)}$ corrects the superposition of
cavity correlation functions for three body interactions. We will hereafter refer to Eq.~\ref{eq:MCA} as the Marshall-Chapman approximation (MCA).

 As shown earlier \cite{bansal_structure_2016}, MCA fails for high densities and high association energies, conditions where
 multi-body interactions are important. But recognizing that the integral in Eq.~\ref{eq:3} is related to $F^{(n)}$ (Eq.~\ref{eq:Fn}) we have  \cite{bansal_structure_2016}
\begin{eqnarray}
\frac{	\Delta c_n^{(0)}}{V}  = {\rho_s^{(0)} {( {x ^{(p)}X_A^{(p)}}f_A^{(s,p)}\sqrt {\kappa _{AA}})}^n} F^{(n)}  \, .
\label{eq:Cn_new}
\end{eqnarray}
It can be observed that all the multi-body correlation information is subsumed in $F^{(n)}$ which is obtained from the occupancy 
distribution $\{p_n\}$. We follow our earlier work \cite{bansal_structure_2016} to estimate this distribution. Importantly, since $\{p_n\}$ forms
the basis of our \textit{complete reference} approach, we also develop an analytical model to describe these distribution functions.

Finally, with the above information, and based on the Marshall-Chapman theory \cite{marshall_thermodynamic_2013}, the fraction of solute associated with $n$ solvent molecules is 
\begin{equation}
X_n^{\left( s \right)} = \frac{ {( {x ^{(p)}X_A^{(p)}}f_A^{(s,p)}\sqrt {\kappa _{AA}})}^n F^{(n)} }{{1 + \sum\limits_{n = 1}^{{n^{\max }}} {( {x ^{(p)}X_A^{(p)}}f_A^{(s,p)}\sqrt {\kappa _{AA}})}^n} F^{(n)} } \, ,
\label{eq:301}
\end{equation}
and the fraction of solute not bonded to any solvent molecule is 
\begin{equation}
X_0^{\left( s \right)} = \frac{1}{{1 + \sum\limits_{n = 1}^{{n^{\max }}} {( {x ^{(p)}X_A^{(p)}}f_A^{(s,p)}\sqrt {\kappa _{AA}})}^n} F^{(n)} } \, .
\label{eq:300}
\end{equation}
Using these distributions for associating mixture, the average number of solvent associated with the solute is given by: 
\begin{equation}
n_{avg} = \sum\limits_n {n\cdot {X^{(s)}_n}}  \, ,
\label{eq:81}
\end{equation}
The fraction of solvent not bonded at site $A$ and site $B$ can be obtained by simultaneous solution of the following equations:
\begin{equation}
X_A^{\left( p \right)} = \frac{1}{{1 + \xi {\kappa _{AB}}f_{AB}^{\left( {p,p} \right)}{\rho_p}X_B^{(p)} + \frac{{{\rho_s}}}{{{\rho_p}}}\frac{{ n_{avg} }}{{X_A^{(p)}}}}}  \, ,
\end{equation}
\begin{equation}
X_B^{\left( p \right)} = \frac{1}{{1 + \xi {\kappa _{AB}}f_{AB}^{\left( {p,p} \right)}{\rho_p}X_A^{(p)}}} \, .
\label{eq:82}
\end{equation}
where   
\begin{eqnarray*}
	\xi  & = & 4\pi {\sigma^2}\left( {{r_c} - \sigma} \right){y_{HS}}(\sigma) \\ 
	\kappa_{AB} & = &\left[1-cos(\theta_c)\right]^2/{4} \\
	f_{AB}^{({p,p})} & = & \exp ( \varepsilon _{AB}^{({p,p})}/k_{\rm B}T)-1 \, .
\end{eqnarray*}

\subsection{Occupancy distribution $\{p_n\}$ for the hard-sphere fluid} \label{sc:HStheory}
Consider a hard sphere fluid with one solute and $N$ solvent particles in a volume $V$ and temperature $T$. We are interested in the occupancy statistics $\{p_n\}$ of the solvent in the coordination volume around the solute. To this end consider the reaction
\begin{equation}
S{P_{n = 0}} + {P_n} \rightleftharpoons S{P_n} \, ,
\label{eq:20}
\end{equation}
with the equilibrium constant  
\begin{equation}
{K_n} = \frac{{{\rho _{S{P_n}}}}}{{{\rho _{S{P_{n = 0}}}}\rho _p^n}} \, ,
\label{eq:21}
\end{equation}
where $\rho_{SP_n}$ is the density of  species $SP_n$ and $\rho_p$ is the density of the solvent. Clearly, we have \cite{pratt_quasichemical_2001,lrp:book,lrp:cpms}
\begin{equation}
\frac{p_n}{p_0} = K_n \rho_p^n \, ,
\label{eq:pnp0}
\end{equation}
where $p_0$ is the probability that the coordination volume is empty of solvent particles.  Following earlier work in studying 
clusters with quasichemical theory \cite{pratt_quasichemical_2001,pratt_selfconsistent_2003,merchant_thermodynamically_2009,merchant:jcp11b}, we can show that
\begin{equation}
K_n = \frac{(e^{ \beta\mu^{\rm ex}_{p} })^n}{n!}\int\limits_v d{\vec r}_1\ldots\int\limits_v d{\vec r}_n e^{-\beta U_{SP_n}(R^n)} e^{-\beta \phi(R^n;\beta)} \, , 
\label{eq:kn1}
\end{equation}
where $U_{SP_n}(R^n)$ is the potential energy of the $n$-solvent cluster (with the solute $S$ fixed at the center of the cluster), $\beta = (k_{\rm B}T)^{-1}$,  $\phi(R^n;\beta)$ is the free energy of interaction of the cluster with the rest of the bulk medium for a given configuration $R^n$ of the cluster, and $\beta\mu^{\rm ex}_{p} $ is the excess chemical potential of the solvent particle. (For completeness, in  appendix~A we derive the above expression for $K_n$.) $\phi(R^n;\beta)$ can also be thought of as a field imposed by the bulk solvent medium \cite{pratt_selfconsistent_2003,merchant:jcp11b} on the solute-solvent cluster in the observation volume. Earlier Pratt and Ashbaugh \cite{pratt_selfconsistent_2003} modeled this field using a self-consistent approach. 
Here we take a different approach. 

First note that without the field term, the cluster integral presents a simpler $n$-body problem (where $n$ is small, typically less than 20 for systems
of interest here). The field is thus an interfacial term that couples the local cluster with the bulk medium. To  make this explicit, 
we can rewrite Eq.~\ref{eq:kn1} as 
\begin{equation}
K_n = \frac{(e^{ \beta\mu^{\rm ex}_{p} })^n}{n!} \langle e^{-\beta \phi(R^n;\beta)}\rangle_0 \int\limits_vd{\vec r}_1\ldots\int\limits_v d{\vec r}_n e^{-\beta U_{SP_n}(R^n)} \, .
\end{equation}
Here $\langle \ldots\rangle_0$ indicates averaging with over the normalized probability density for cluster conformations $R^n$ in the absence of interactions with the rest of the medium, i.e.\ over the density $e^{-\beta U_{SP_n}(R^n)}  / (n! K_n^{(0)})$, where 
\begin{equation}
n! K_n^{(0)} =  \int\limits_vd{\vec r}_1\ldots\int\limits_v d{\vec r}_n e^{-\beta U_{SP_n}(R^n)} \, , 
\label{eq:kn0}
\end{equation}
and the interfacial contribution is $\beta\Omega_n = -\ln \langle e^{-\beta \phi(R^n;\beta)}\rangle_0$. 

From analysis of simulation data for different densities, we find that $\Omega_n$ can be described by a two parameter equation as 
\begin{equation}
\beta \Omega_n =-\zeta_1 \cdot  n^2+\zeta_2\cdot n \, .
\end{equation}
This model of the interfacial term was anticipated in our previous work \cite{bansal_structure_2016}. Thus we finally obtain 
\begin{equation}
{p_n} = \frac{\exp(\beta \cdot n \cdot \mu^{ex}_{p})\rho_p^n [{\exp{(\zeta_1 \cdot  n^2-\zeta_2\cdot n)}}]{K_n}^{(0)}}{{1 + \sum\limits_{j \ge 1} {\exp(\beta \cdot j \cdot \mu^{ex}_{p})\rho_p^j [{\exp{(\zeta_1 \cdot j^2-\zeta_2 \cdot j)}}]{K_j}^{(0)}} }} \ 
\label{eq:pn_2par}
\end{equation}
The above equation can also be derived using a two moment maximum entropy approach,  with the mean and variance of the occupancy as constraints and $K_n^{(0)}$ as the default (see appendix~A). 

Drawing upon the work of Reiss and Merry  \cite{reiss_upper_1981}, we can model the interfacial term in terms of surface sites (of the cluster) 
that are available to interact with the bulk fluid. On the basis of such a mean field approach and guided by Monte Carlo (MC) simulation data for different densities of hard sphere systems, we find that 
\begin{equation}
\beta \Omega_n =(-0.0109 \cdot  n^2+1.0109 \cdot n)\cdot \zeta
\label{eq:surf_int}
\end{equation}
\begin{equation}
{p_n} = \frac{\exp(\beta \cdot n \cdot \mu^{ex}_{p})\rho_p^n [{\exp{(-(-0.0109 \cdot  n^2+1.0109 \cdot n)\cdot \zeta)}}]{K_n}^{(0)}}{{1 + \sum\limits_{j \ge 1} {\exp(\beta \cdot j \cdot \mu^{ex}_{p})\rho_p^j [{\exp{(-(-0.0109 \cdot  j^2+1.0109 \cdot j)\cdot \zeta)}}]{K_j}^{(0)}} }} \ 
\label{eq:pn_1par}
\end{equation}

Eq.~\ref{eq:pn_2par} and Eq.~\ref{eq:pn_1par} are the 2-parameter and 1-parameter models, respectively, for $p_n$ on the basis of which 
we obtain $F^{(n)}$ (Eq.~\ref{eq:Fn}) to describe multi-body correlations in the reference fluid.  The parameter values for different densities are given in Table \ref{table: param1}. These parameters were obtained based on hard spheres mixtures with all particles of the same size. Since the
information about size-asymmetry is already contained in the isolated cluster partition function, we will use these parameters to study asymmetric mixtures and will discuss limitations for cases with extreme size ratio.
    
    \section{METHODS}
       
       To compare the theory results, we perform Monte-Carlo (MC) simulations for a range of systems. This section presents the details of the MC simulations for different associating and hard sphere systems.The  Marshall-Chapman approximation (MCA) and the models developed for hard sphere distribution functions require isolated cluster probabilities; these were also computed for different size ratios. 
   \subsection{Monte Carlo Simulations}

MC simulations were carried out for reference hard sphere systems and associating systems to compare the results of Marshall-Chapman theory using MCA and the complete reference approach. The associating mixture contains the patchy solvent particles and the 
isotropically interacting solute defined by the potentials given by Eq.~\ref{eq:potential1} and Eq.~\ref{eq:potential2}, respectively. 
The solute diameter is $\sigma_s$ and solvent diameter is $\sigma_p$. 
    
The observation volume is defined by a critical radius $r_c = 1.1\bar{\sigma}$,  where $\bar{\sigma} = (\sigma_s + \sigma_p)/2$ is the closest distance of approach. For cases where $\sigma_s/\sigma_p \geq 1.5$, a cutoff of 1.1$\bar{\sigma}$ can include some of the second-shell solvent. To avoid this and focus attention to the first observation shell, we set $r_c = \bar{\sigma} + 0.1\sigma_p$ for these cases. In the associating system (Fig.~\ref{fig:1} and Eqs.~\ref{eq:potential1} and~\ref{eq:potential2}), the critical angles for interaction are $\theta_c^{(A)} = \theta_c^{(B)} = 27\degree$. 
    
    For hard sphere mixtures 255 solvent particles and 1 solute particle were studied in a given simulation cell. Ensemble reweighting technique was used to reveal low probability states \cite{merchant_water_2011}. The system was equilibrated for 1 million steps with translational factors chosen to yield an acceptance rate of 0.3, and data was collected every 100 sweeps. Analysis was carried out for different densities and size ratios $\sigma_s/\sigma_p$.
    
    For associating mixtures bonding distributions and average bonding numbers were studied for mixtures with different sizes and different association strengths for solute-solvent and solvent-solvent interactions. The excess chemical potential of the coupling of the colloid with solvent was also calculated using thermodynamic integration of average binding energy of solute with solvent as a function of solute-solvent interactions, using the three-point Gauss Legendre quadrature technique \cite{Hummer:jcp96}. For a symmetric mixture with no solvent-solvent interactions, energetic and entropic contributions for solute chemical potential were also studied at constant volume and temperature.
   
   Concentration effects were also computed considering a total of 864 particles, with varied number of solute particles. Due to the difference in size of the solute and solvent, the computations were performed keeping the packing fraction constant. Hence, the density of the system changed with respect to the concentration. Also the maximum angle for which the patch can form single bond is computed using the law of cosines, and hence the critical angle needs to be altered when the solute size is smaller than the solvent. For a size ratio of $\sigma_s/\sigma_p=0.8$, critical angle $\theta_c^{(A)} = \theta_c^{(B)} = 20\degree$ was used to ensure single bonding condition for the $A$ patch on solvent molecules which can associate with solute molecules.
    
    \subsection{Isolated cluster probabilities}
    
   For asymmetric mixtures, we also study isolated cluster probabilities for different size ratios of solute and solvent molecules. The observation volume around the isolated solute is the same as defined in the previous section. For different size ratios we use the spherical code (appendix~B) to estimate the maximum number of solvent molecules which can be inserted in the observation volume of the isolated solute. Successive insertion probabilities are calculated as in previous works \cite{marshall_molecular_2013,bansal_structure_2016}, where $10^8$ to $10^9$ insertions were carried out in the observation volume around the solute and the cases with no overlap with remaining $n-1$ particles studied. The data for isolated cluster probabilities for different sizes is given in the appendix~B.

\section{Results and discussions} 
\subsection{Hard sphere $\{p_n\}$ distribution}\label{sc:res_hs}
\begin{figure*}[ht!]
	\centering
	\includegraphics[scale=1]{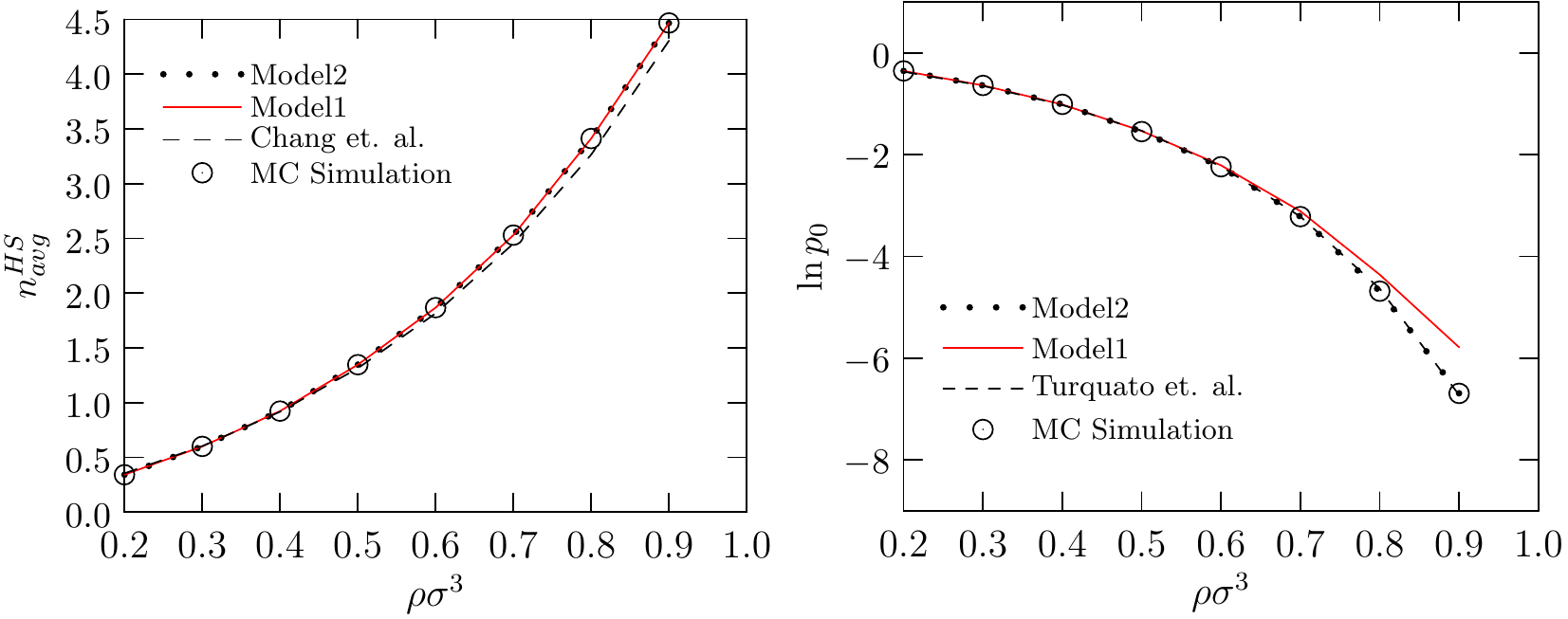}
	\caption{Average occupancy $n^{HS}_{avg}$ and exclusion probability $\ln p_0$ for symmetric hard sphere mixture. Model1 (Eq.~\ref{eq:pn_1par}) and Model2 (Eq.~\ref{eq:pn_2par}) correspond to the 1 and 2-parameter models, respectively. 
Chang et. al. corresponds to the study in Ref.~\cite{chang_real_1994} for $n^{HS}_{avg}$ based on Percus-Yevick approximation. Torquato et. al. corresponds to the study in Ref.~\cite{torquato_microstructure_1985} for exclusion probabilities based on Carnahan-Starling approximation.}
	\label{fig:HS_sym} 
\end{figure*}

Recall that Eq.~\ref{eq:pn_1par} and Eq.~\ref{eq:pn_2par} are the 1-parameter and 2-parameter models for the occupancy 
distribution $\{p_n\}$. To obtain the parameters for hard spheres all of the same size, for both the models we use 
the average occupancy ($n^{HS}_{avg}= \sum\limits_n n\cdot {p_n}$) as a fitting constraint. For the 2-parameter model
we additionally use the exclusion probability ($p_0$) --- the probability when no hard sphere solvent particle
is present in the observation volume --- as a constraint. For the 1-parameter model we study the surface interactions based on the mean field approach developed by Reiss and Merry  \cite{reiss_upper_1981}. By analyzing the distribution functions $\{p_n\}$ for different densities, we obtain geometric effects (density independent) that describe the mutual interference of different surface sites (Eq.~\ref{eq:surf_int}). The density (or packing fraction) dependent parameters for these two models are given in Table ~\ref{table: param1}. 
\begin{table}[ht]
	\caption{Parameters for Eq.~\ref{eq:pn_2par} ($\zeta_1$,$\zeta_2$) and Eq.~\ref{eq:pn_1par} ($\zeta$) for different densities $\rho \sigma^3$ and packing fractions ($\eta$) }  
	\begin{tabular}{|c|c| c| c| c|}
		\hline
		$\rho \sigma^3$ & $\eta$  & $\zeta_1$  & $\zeta_2$& $\zeta$  \\
		\hline
0.2	&	0.105	&	0.0175	&	0.773	&	0.75	\\
0.3	&	0.157	&	0.015	&	1.267	&	1.251	\\
0.4	&	0.209	&	0.0256	&	1.979	&	1.947	\\
0.5	&	0.262	&	0.023	&	2.88	&	2.876	\\
0.6	&	0.314	&	0.0361	&	4.179	&	4.172	\\
0.7	&	0.367	&	0.0457	&	5.947	&	5.985	\\
0.8	&	0.419	&	0.0609	&	8.432	&	8.562	\\
0.9	&	0.471	&	0.0829	&	12.088	&	12.414	\\
		\hline
	\end{tabular}
\label{table: param1}
\end{table}  
Fig.~\ref{fig:HS_sym} presents the  results corresponding to the average occupancy ($n^{HS}_{avg}$) and exclusion probability ($\ln p_0$) based on the models. We compare these results with the MC simulation values and also include results 
from literature \cite{chang_real_1994,torquato_microstructure_1985}. 

Fig.~\ref{fig:HS_sym} makes it clear that 2-parameter model can simultaneously capture both $n_{avg}$ and $\ln p_0$ 
in excellent agreement with simulation. Importantly, even the 1-parameter model is able to capture most of the details, affirming the physical ideas underlying the models (Eqs.~\ref{eq:pn_1par} and ~\ref{eq:pn_2par}). 

\begin{figure*}[ht!]
	\centering
	\includegraphics[scale=1]{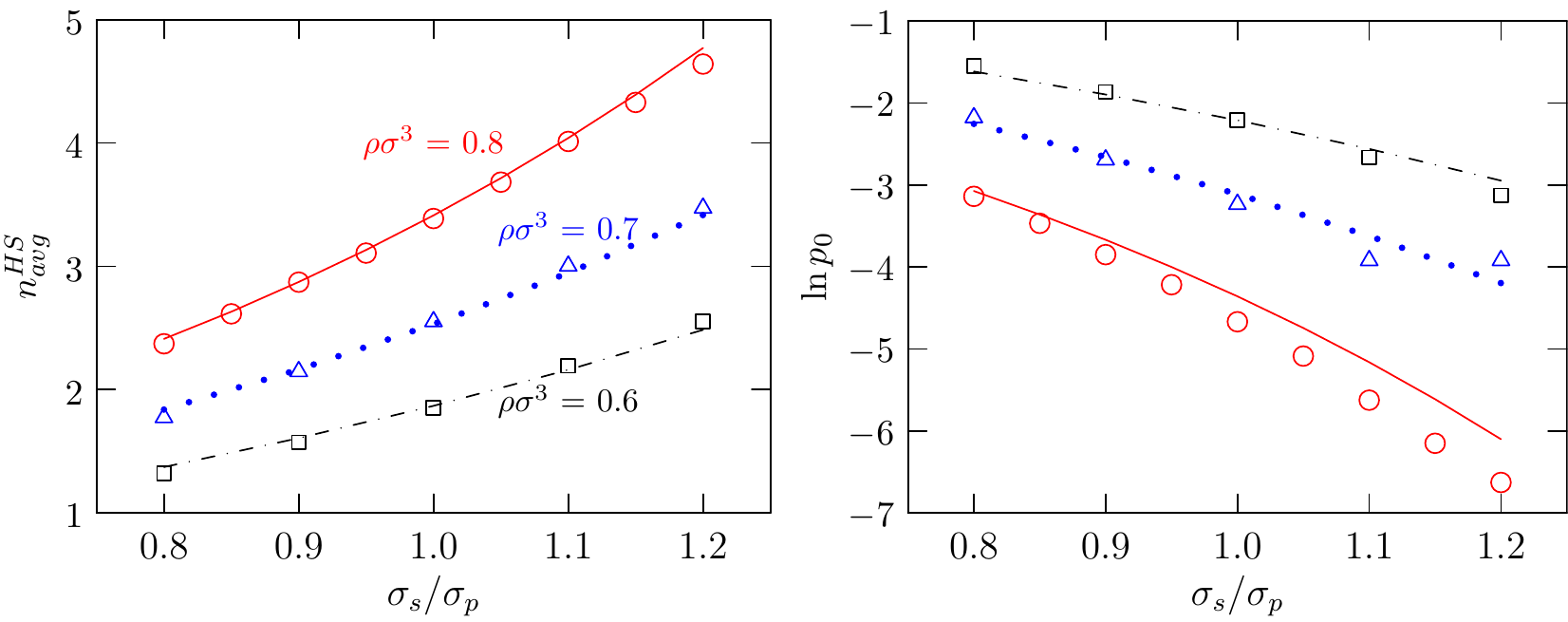}
	\caption{Average occupancy ($n^{HS}_{avg}$)  and exclusion probability ($\ln p_0$) for different size ratios and different reduced densities. Symbols represent MC simulation values and lines represent the prediction results based on 1-parameter model (Eq.~\ref{eq:pn_1par}).}
	\label{fig:pn1_asym} 
\end{figure*}

Next, using the parameters obtained above for a fluid where  both solute and solvent hard-spheres are the same size (a symmetric mixture), we describe the occupancy in a fluid where the solute and solvent are of different sizes (an asymmetric mixture). Our \textit{ansatz} is that 
the information about size asymmetry is adequately captured by the isolated cluster partition function $K_n^{(0)}$ (Eq.~\ref{eq:kn0}). 
For an infinitely dilute system comprising one solute in a solvent bath, Fig.~\ref{fig:pn1_asym} shows the predictions of $n_{avg}^{HS}$ and $\ln p_0$ based on the 1-parameter model (Eq.~\ref{eq:pn_1par}) for different size ratios and different reduced densities. The level of agreement is encouraging, but perhaps not surprising since the size ratios are not much different from a symmetric case. Thus the geometric effects describing the mutual interference of different surface sites in the packing around the
solute will be similar to what is observed in the symmetric mixture.  For extreme size ratios this should break down, as we will 
discuss in the context of average bonding numbers in associating mixtures.  

\begin{figure*}[ht!]
	\centering
	\includegraphics[scale=1]{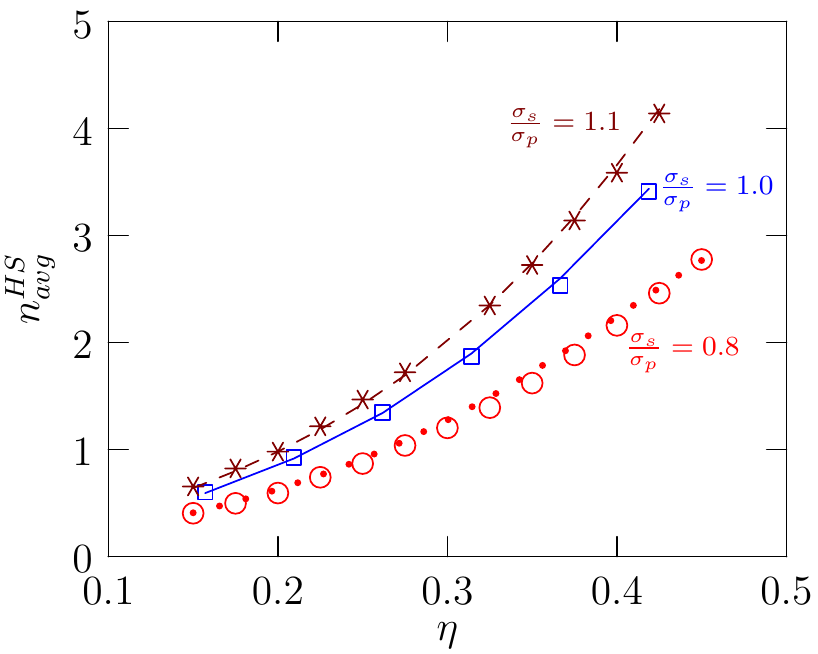}
	\caption{Variation of average occupancy ($n^{HS}_{avg}$) for different size ratios with packing fraction. Symbols represent MC simulation values and curves represent the prediction results based on 1-parameter model (Eq.~\ref{eq:pn_1par}). Solute is infinitely dilute.}
	\label{fig:pn1_asym_eta} 
\end{figure*}

For asymmetric mixtures, the packing fraction is a better measure of packing in the fluid. Fig.~\ref{fig:pn1_asym_eta} 
presents the variation of average occupancy for different packing fractions for three different size ratios for an infinitely dilute solution. The results show that the 1-parameter model is able to predict the average occupancy quite well. We note that as the concentration of solute is changed in asymmetric mixtures, the packing fraction changes (for a given density) and parameters corresponding to the 
resulting packing fraction should be used from Table~\ref{table: param1}.

\subsection{Associating mixture}\label{sc:res_asso}

We next consider associating fluids and investigate both size and concentration effects. 

\subsubsection{Infinite Dilution}
\begin{figure*}[ht!]
	\centering
	\includegraphics[scale=1]{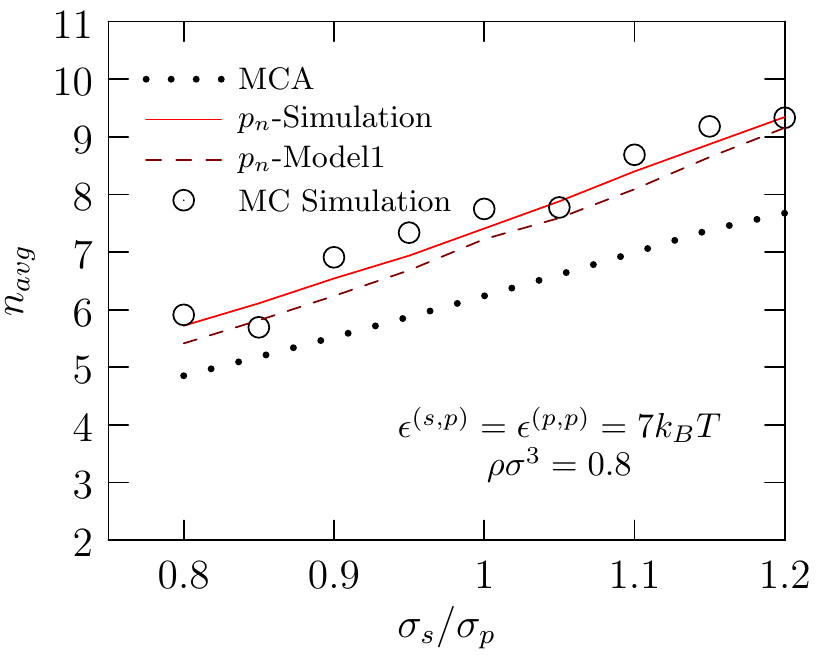}
	\caption{Average bonding numbers $n_{avg}$ in associated mixtures for asymmetric cases with different relative size of solute molecules. Solute is infinitely dilute and density is $\rho\sigma^3=0.8$. Symbols represent MC simulation values. `$p_n$-Simulation' corresponds to the complete reference theory results with $\{p_n\}$ distribution computed directly from simulations for calculating $F^{(n)}$ (Eq.~\ref{eq:Fn}). `$p_n$-Model1' corresponds to the use of 1-parameter model to compute $\{p_n\}$ in our theory. `MCA' corresponds to the second order perturbation approximation used by Marshall-Chapman (Eq.~\ref{eq:MCA}).}   
	\label{fig:n_avg} 
\end{figure*}

We first study an infinitely dilute solution and vary the size of solute with respect to a fixed size of solvent particles.  
In our complete reference theory, the reference fluid $\{p_n\}$ distribution is either computed directly from simulations (`$p_n$-Simulation' in Fig.~\ref{fig:n_avg}) or from the 1-parameter model discussed in Sec.~A above (`$p_n$-Model1' in Fig.~\ref{fig:n_avg}). Using $\{p_n\}$ we compute $F^{(n)}$ (Eq.~\ref{eq:Fn}), and on that basis, the average number of bonds in the associating mixture using Eq.~\ref{eq:Cn_new}, Eq.~\ref{eq:301} and Eq.~\ref{eq:81}. Fig.~\ref{fig:n_avg} shows the variation of average bonding numbers with size ratio of solute and solvent molecules for a density of 0.8 and association strength of 7 $k_{\rm B}T$. 

As the size of the solute increases with respect to the size of the solvent, more solvent molecules can associate with the solute. With accurate information about the reference fluid $\{p_n\}$ (and hence $F^{(n)}$) from direct simulations, the complete reference theory is able to capture this increase in bonding numbers quite accurately. Interestingly, even $\{p_n\}$ obtained using the 1-parameter model (Eq.~\ref{eq:pn_1par}) suffices. But note that the Marshall-Chapman approximation with only up to 3-body effects incorporated in the theory
underestimates the average bonding numbers. These results emphasize the importance of multi-body interactions in describing the association correctly.

As noted in Sec.~ \ref{sc:HStheory}, the amount of surface exposure (or surface sites)  is an important factor in determining the packing effects in the models. 
For the size ratios considered in Fig.~\ref{fig:n_avg}, the maximum number of surface sites and hence the geometric effects in surface interactions are expected to be similar to the symmetric case,
and not surprisingly, the agreement of bonding numbers between simulation and with the 1-parameter model for $p_n$ is very good. For extreme size ratios (Table.~\ref{table:extreme}), a high error with the 1-parameter model is expected. This results because of the disparity in surface sites for these ratios relative to the symmetric mixture within which the density independent geometric effects were obtained (Eq.~\ref{eq:surf_int}). However, 
even for these extreme size ratios, with information for reference hard sphere distribution functions from simulation, the complete reference theory is able to capture the average bonding numbers for these extreme size ratios quite well.  
 \begin{table}[ht]
 	\caption{Comparison of average bonding numbers ($n_{avg}$) of solute for extreme size ratios.}  
 	
 	\begin{tabular}{|c| c| c| c|}
 		\hline
 	$\sigma_s / \sigma_p $ & MC & $p_n$-Simulation & $p_n$-Model1 \\
 		\hline 		
 	0.5	&	3.61	& 3.55	& 4.50 	\\
 	2	&	15.84	& 15.04 & 17.66	\\ 		
 		\hline
 	\end{tabular}
 	\label{table:extreme}
 \end{table}

\subsubsection{Varying association strengths}
\begin{figure*}[h!]
	\centering
	\includegraphics[scale=1]{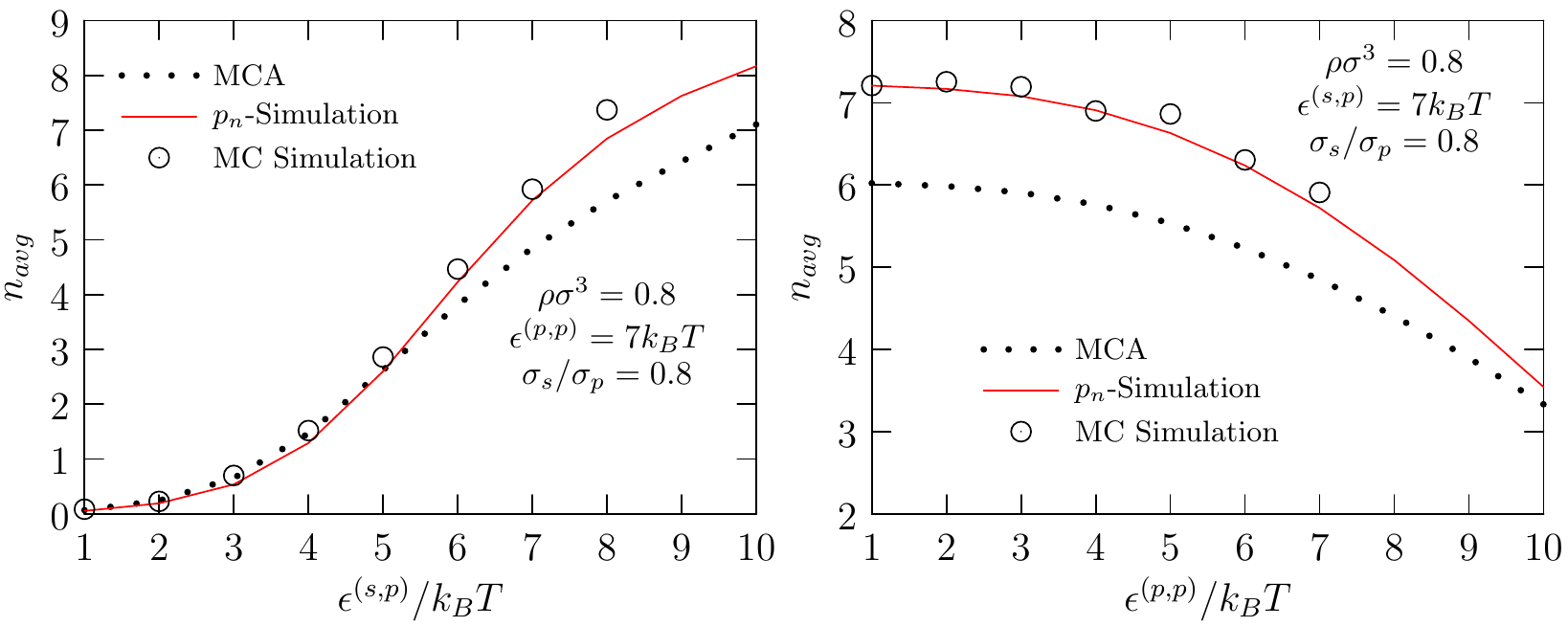}
	\caption{Variation of average bonding numbers ($n_{avg}$) with the relative strengths of solute-solvent and solvent-solvent association strengths. Left: solute-solvent association strength is varied with solvent-solvent association fixed at $\epsilon^{(p,p)}=7k_{\rm B}T$. Right: solvent-solvent association strength is varied with solute-solvent association fixed at $\epsilon^{(s,p)}=7k_{\rm B}T$. Solute is infinitely dilute and density is $\rho\sigma^3=0.8$. Rest of the representation is the same as Fig.~\ref{fig:n_avg}. }
	\label{fig:Asso_asym} 
\end{figure*}

 To understand the effect of varying association strengths between solute-solvent and solvent solvent particles, we studied a
 case with a fixed size ratio of $\sigma_s / \sigma_p = 0.8 $ and for varying association strengths (fig.~\ref{fig:Asso_asym}). 

As the strength of solute-solvent association is increased as compared to solvent-solvent interactions, multi-body effects become important. It was observed that higher deviations are observed with TPT2-based Marshall-Chapman approximation, 
especially for increasing strength of solute-solvent association. Importantly, excellent agreement with the 
complete reference theory is observed for all cases noted in the figure. 

\subsubsection{Chemical potential and energy entropy contributions}

\begin{figure*}[h!]
	\centering
	\includegraphics[scale=1]{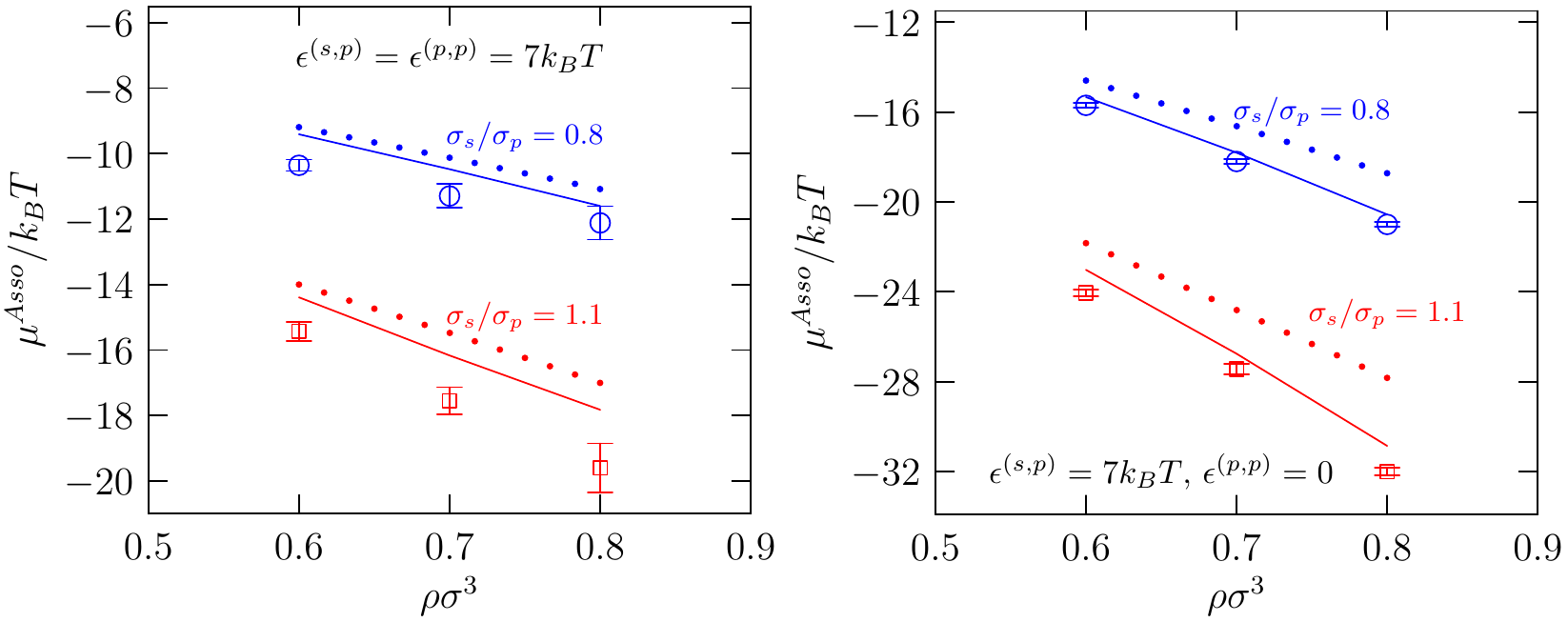}
	\caption{Chemical potential for charging a solute molecule in a patchy solvent environment for different reduced densities and different size ratios. The solute is infinitely dilute and association energy for interactions between solute-solvent molecules is $7k_{\rm B}T$. Solvent-solvent interaction energy is 7$k_{\rm B}T$(left) and zero (right). Symbols represent simulation values, solid curves are complete reference theory results with $p_n$ from simulations and dotted curves corresponds to results with 'MCA'.}
	\label{fig:Asso_mu} 
\end{figure*}

Fig.~\ref{fig:Asso_mu} shows the chemical potential of the solute for two limiting cases: one where the solvent-solvent association is 
present and one where it is absent, for two size ratios. 
The results indicate that despite the large deviations in $n_{avg}$ (Fig.~\ref{fig:n_avg}) noted above,  the deviations using the Marshall-Chapman approximation are not as high relative to the complete reference approach. 

To better understand this result, we  consider a symmetric system where there is no association between solvent particles. For this case, the partial molar energy is the energy of the system with solute-solvent association and there is no contribution for change in solvent-solvent association due to inclusion of an ideal solute at infinite dilution. We decompose the
residual chemical potential of the solute ($\mu^{Asso}_{s}$) into its energetic and entropic contributions  \cite{yu_thermodynamic_1988,yu_solvation_1990}: 
 
  \begin{equation}
  \beta \mu_s^{Asso} =  \beta E_s^{Asso}-T\cdot \beta S_{s,V}^{Asso}
  \end{equation}
  where $ E_s^{Asso} =  ( \frac{\partial \mu^{Asso}_s/T}{\partial 1/T})_{{\rho _s},{\rho _p}}$ is the partial molar energy and $ S_{s,V}^{Asso}=  {( {\frac{{\partial \mu _s^{Asso}}}{{\partial T}}})_{{\rho _s},{\rho _p}}}$ is the partial molar entropy contribution.  The entropy and energy values can be obtained with MCA and complete reference approach based on the corresponding temperature derivatives of the chemical potential of solute. 
\begin{figure*}[ht!]
	\centering
	\includegraphics[scale=0.65]{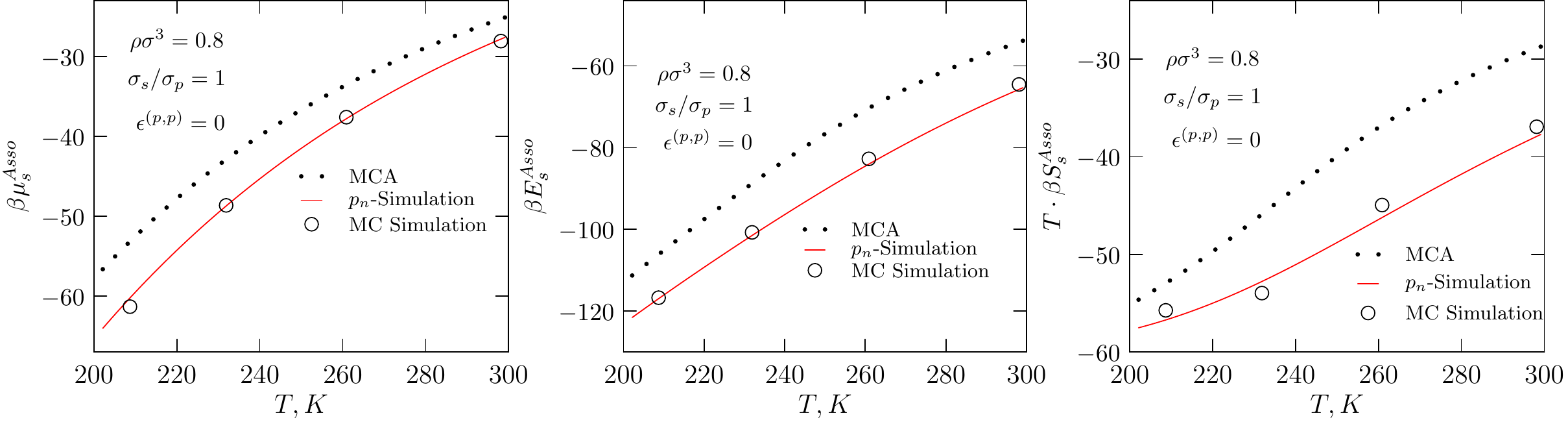}
	\caption{Variation of residual chemical potential of solute($\mu_s^{Asso}$), partial molar energy $(E_s^{Asso})$ and entropy ($S_{s,V}^{Asso}$) with temperature for a system with no solvent-solvent association interactions. The interactions between solute-solvent for different temperatures is determined by $\frac {\epsilon^{(s,p)}}{k_{\rm B} T }=2087.05/T $. }
	\label{fig:en_entr} 
\end{figure*}

Fig.~\ref{fig:en_entr} clearly shows that energy and entropy contributions are not captured accurately by MCA. But the apparent
reasonable prediction of the chemical potential (using MCA) results because of cancellation of the errors between the entropy and energy contributions. The above result shows that in comparing perturbation theories, it could be useful and prudent to study chemical potential and also 
its energy and entropy contributions. 
\subsubsection{Concentrated systems}
\begin{figure*}[ht!]
	\centering
	\includegraphics[scale=1]{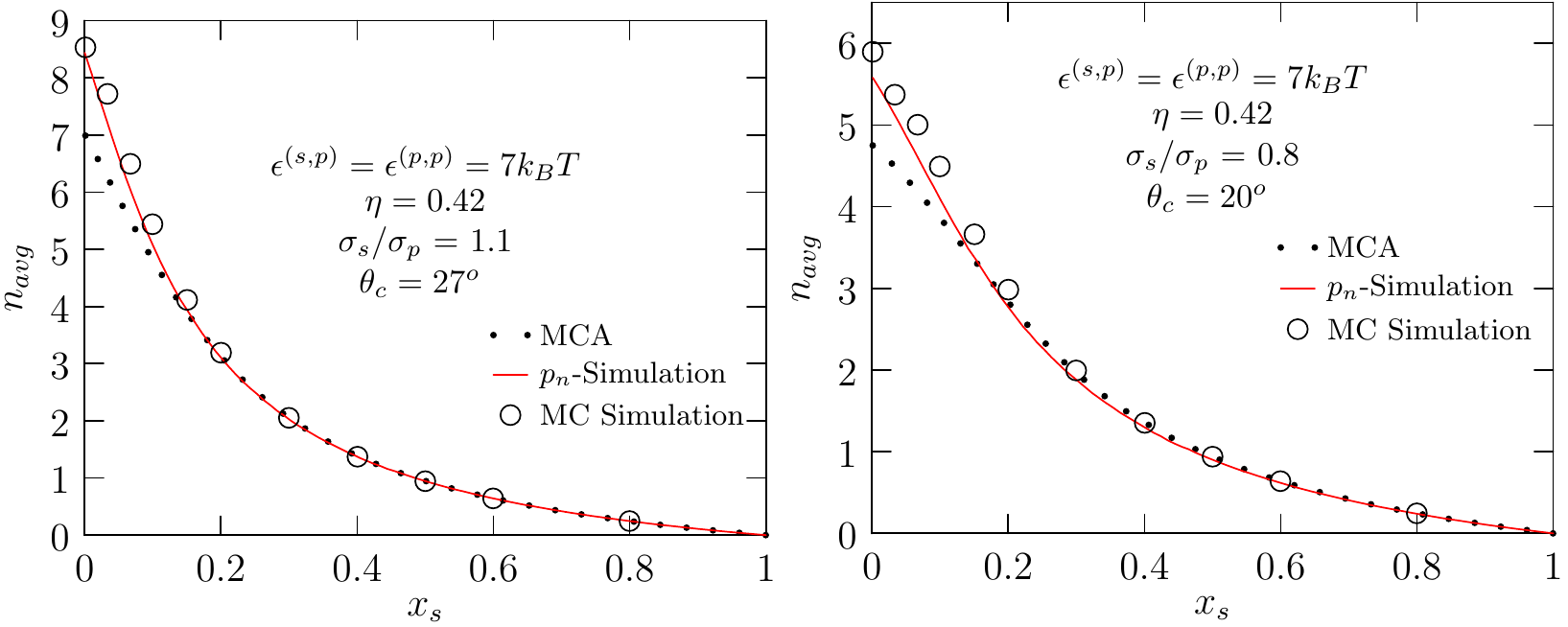}
	\caption{Variation of average bonding number of solute ($n_{avg}$) with the concentration of solute for size ratio, $\sigma_s/\sigma_p$, 1.1 (left) and 0.8 (right). Note that to ensure the single bonding condition for solvent sites, the critical angle has been reduced to $\theta_c=20^o$ for 0.8 size ratio. }
	
	\label{fig:n_avg_conc} 
\end{figure*}

We also study the variation of average bonding number of the solute for different concentration of the solute in the asymmetric mixtures. As the concentration of the solute increases, the system becomes limited in the number of solvent molecules that can bond to the solute molecules and hence TPT2 approximation used in MCA becomes more accurate. For low concentrations, multi-body correlations are important and hence deviations are observed with MCA. Fig.~\ref{fig:n_avg_conc} shows the results for two different size ratios. The theory with the complete reference is able to capture the average bonding numbers for the whole concentration range. 

\section{Concluding discussions}

We have studied asymmetric mixtures having strong short-range association between differently sized solute and solvent molecules. The solute molecules have isotropic interactions and the solvent molecules have directional interactions. Such systems are archetypes of colloidal mixtures that are being actively studied in designing materials from the nanoscale level. These systems can also describe the short range ion-solvation and ion-pairing effects in electrolyte systems which is another focus of our research.

The isotropic interactions of the solute can allow multiple solvent molecules to associate and hence multi-body effects become important for these systems. Previously \cite{bansal_structure_2016}, we discussed the development of an accurate perturbation theory for these systems is hindered by the difficulty in obtaining the multi-body correlations in the reference system (typically hard sphere). We discussed the limitation of an approach based on obtaining multi-body correlations for the hard sphere system in the gas phase and approximating bulk solvent effect with a linear superposition of pair correlation (together with term to account for three body correction). It was observed that this second order perturbation method 
fails at high association strengths and high densities. We introduced an approach to represent the multi-body clusters in terms of occupancy distribution to accurately describe the packing in the hard sphere system. Excellent agreement with MC simulation for a range of conditions of association and concentrations were obtained with this \textit{complete reference} approach for symmetric mixtures. 

Here, we have built upon the earlier work and study systems with size asymmetry. Our study shows that the multi-body correlations for asymmetric hard sphere mixtures can be accurately studied in terms of occupancy distributions. With these accurate packing effects, our approach gives excellent agreement with MC simulation for the asymmetric associating mixtures. These occupancy distributions were obtained by particle simulations. Based on ideas borrowed from quasichemical theory, we have also developed parametric models to describe occupancy distributions in the hard sphere systems of different densities and different asymmetries. These distributions were obtained by describing the effects of clustering, medium and surface interactions simultaneously in the hard sphere packing around a solute and can be incorporated in perturbation theories (for eg. statistical associating fluid theory) without having to perform particle simulations for the reference fluid. We validate this complete reference theory (using parameterized
models of the reference distribution) against several simulations. A critical test was in analyzing the energy and entropy contributions to the chemical potential of the solute. For a system where solvent-solvent association is not present, such a decomposition of chemical potential in energy and entropy contributions with {complete reference} theory showed excellent agreement with MC simulations. The apparently reasonable agreement of the
second order perturbation approach is shown to arise from the balancing of errors in the energy and entropy contributions. This important finding
suggests the need to study different properties while validating perturbative theories for fluids. This present framework can prove useful in modeling real solutions where concentration of solute is low and its size is different from the solvent molecules. 
 \section{Acknowledgment} 
   We acknowledge RPSEA / DOE 10121-4204-01 and the Robert A. Welch Foundation (C-1241) for financial support.  

\newpage
\section{Appendix}\label{sc:appen}
\subsection{Hard sphere distribution}
\subsubsection{Expression for $K_n$}\label{sc:Kn}
From Eq.~\ref{eq:pnp0}, we find that obtaining an expression for $K_n$ reduces to evaluating the ratio of $p_n / p_0$. The total potential energy of the system when $n$ solvent particles are coordinated with the solute and the remaining $N-n$ solvent particles are outside the observation volume
can be formally written as $U = U_{SP_n} + U_{N-n|SP_n} + U_{N-n}$. $U_{SP_n}$ is the potential energy of the solute-$n$-solvent cluster.
$U_{N-n|SP_n}$ is the interaction energy of the cluster with the rest of the solvent; specifically $U_{N|SP_0}$ is the interaction energy of the solute
with the fluid outside the observation volume. In the particular case of hard-spheres, $U_{N|SP_0} = 0$. Finally, $U_{N-n}$ is the potential energy of the solvent constituting the bulk. 

Since $p_n \propto Q(n,N-n,V,T)$, where $Q(n,N-n,V,T)$ is the canonical partition function of the system with $n$ solvent in the observation volume around the solute and $N-n$ in the bulk, we immediately have
\begin{equation}
\frac{p_n}{p_0} = \frac{N!}{(N-n)!n!} \frac {\int\limits_v d{\vec r}_1\ldots \int\limits_v d{\vec r}_n e^{-\beta U_{SP_n}} \cdot \int\limits_{V-v} d{\vec r}_{n+1} \ldots \int\limits_{V-v} d{\vec r}_{N-n} e^{-\beta U_{N-n|SP_n}} e^{-\beta U_{N-n}}}{\int\limits_{V-v} d{\vec r}_1 \ldots \int\limits_{V-v} d{\vec r}_N e^{-\beta U_{N}}} \, 
\label{eq:pnporatio}
\end{equation}
where we have implicitly moved the center of the coordinates to the center of the solute and thus canceled a common factor of $V$ from both the numerator and denominator. Further, since $U_{SP_0} = 0$ and $U_{N|SP_0} = 0$, the denominator simply depends on the potential energy of the solvent in the bulk. (Of course for a general solute, this restriction is easily removed \cite{merchant_thermodynamically_2009,merchant:jcp11b}.)

Next consider the ratio
\begin{equation}
\frac{Q(0,N,V-v)}{Q(0,N-n,V-v)} = \frac{Q(0,N-n+1,V-v)}{Q(0,N-n,V-v)}\cdot \frac{Q(0,N-n+2,V-v)}{Q(0,N-n+1,V-v)}\ldots \frac{Q(0,N,V-v)}{Q(0,N-1,V-v)}
\label{eq:pdt}
\end{equation}
where we suppress $T$ for conciseness and the 0 indicates that there is no solute in the system (or as is the case here, $U_{SP_0} = U_{N|SP_0} = 0$).
In the thermodynamic limit of large $V >> v$ and $N >> n$, from the standard potential distribution relation \cite{lrp:book,lrp:cpms,widom:jpc82}, each of the above factor on the right is simply $e^{-\beta \mu_p^{\rm ex}} / \Lambda_p^3 \rho_p$, where $\Lambda_p$ is the thermal de Broglie wavelength of the solvent
sphere and $\mu^{\rm ex}_p$ is its excess chemical potential, and $\rho_p$ the density of solvent. 

Since, 
\begin{equation}
Q(0,N-n,V-v,T) = \frac{1}{\Lambda_p^{3(N-n)} (N-n)!} \int\limits_{V-v}d{\vec r}_1\ldots \int\limits_{V-v} d{\vec r}_{N-n} e^{-\beta U_{N-n}} \,
\end{equation}
we multiply and divide Eq.~\ref{eq:pnporatio} by the factor $\int\limits_{V-v}d{\vec r}_1\ldots \int\limits_{V-v} d{\vec r}_{N-n} e^{-\beta U_{N-n}}$. 
Rearranging the resulting equation using  Eq.~\ref{eq:pdt} in the large $V$ and large $N$ limit, and noting that the 
momentum partition functions (for both solute and solvent) cancel exactly, we obtain Eq.~20 (main text), where 
\begin{eqnarray}
e^{-\beta \phi(R^n; \beta)} & = &  \frac{ \int\limits_{V-v} d{\vec r}_{n+1} \ldots \int\limits_{V-v} d{\vec r}_{N-n} e^{-\beta U_{N-n|SP_n}} e^{-\beta U_{N-n}}}{\int\limits_{V-v}d{\vec r}_1\ldots \int\limits_{V-v} d{\vec r}_{N-n} e^{-\beta U_{N-n}}} \nonumber \\
& = & \langle e^{-\beta U_{N-n|SP_n}} \rangle_{N-n} \,
\end{eqnarray}
where $ \langle \ldots \rangle_{N-n}$ denotes averaging over the configurations of the $N-n$ solvent particles in the volume outside the observation shell. 

\subsubsection{MaxEnt model for $\{p_n\}$}\label{sc:MaxEnt}

Here we present an alternative derivation of the two parameter model (Eq.~\ref{eq:pn_2par}) on the basis of information theoretic modeling
of $\{p_n\}$ \cite{sivia,lrp:jpcb98,asthagiri:pre03}. On the basis of the isolated cluster partition function, 
we have the distribution of occupancy probabilities $\{p_n^{(0)}\}$ as
\begin{equation}
p_n^{(0)} = \frac{K_n^{(0)} \rho_p^{n}}{1+\sum\limits_{m\ge 1} K_m^{(0)} \rho_p^{m}}
\end{equation}

With $\{p_n^{(0)}\}$ as the default model and accepting the availability of the mean (first moment) and variance (second moment) of the distribution $\{p_n\}$ from simulation data, by standard maximum entropy arguments, we have 
\begin{equation} \label{it04}
	\frac{p_n}{p^0_n} = e^{-C} e^{-\lambda_1 \cdot n} e^{-\lambda_2 \cdot n^2} \, 
\end{equation}
where the Lagrange multipliers $C$, $\lambda_1$, and $\lambda_2$ are, respectively, obtained from enforcing the following constraints 
\begin{equation} \label{it05}
	\sum_n p_n = 1
\end{equation}
\begin{equation} \label{it06}
  \sum_n p_n \cdot n = n_{avg}
\end{equation}
\begin{equation} \label{it07}
	\sum_n p_n \cdot n^2= -\overline{\sigma^2}
\end{equation}

We thus obtain
\begin{equation} \label{it11}
	p_n = \frac{\big [ e^{-\lambda_1} e^{-\lambda_2 \cdot n} \big ]^{n} K_n^{(0)} \rho_p^{n}}{1+ \sum \limits_{m\ge 1} \big [ e^{-\lambda_1} e^{-\lambda_2 \cdot m} \big ]^{m} K_m^{(0)} \rho_p^{m}}
\end{equation}
Eq.~\ref{it11} is the same form as obtained in section \ref{sc:HStheory} for a two parameter correlation. By using $\lambda' = e^{-\lambda_1}$, we can also represent Eq.~\ref{it11} by

\begin{equation} \label{it13}
	p_n = \frac{\big [ \lambda' e^{-\lambda_2 \cdot n} \big ]^{n} K_n^{(0)} \rho_p^{n}}{1+ \sum\limits_{m\ge 1} \big [\lambda' e^{-\lambda_2 \cdot m} \big ]^{m} K_m^{(0)} \rho_p^{m}}
\end{equation}
which is the same form introduced in our previous work \cite{bansal_structure_2016}. Based on this derivation, we can see that $ \lambda_2 $ is the term corresponding to surface interactions discussed in section ~\ref{sc:HStheory} and constraints the variance of the $\{p_n\}$ distribution.
 \subsection{Isolated cluster probabilities}  
 For asymmetric mixtures we study the isolated cluster probabilities and to find the maximum number of solvent molecules that can occupy the observation volume, we use spherical code \cite{sphcode,sph_code2}.
 \subsubsection{Spherical Code}
 The spherical code provides information to place $n$ points optimally across a sphere. It gives the optimal angle between the points, considering the center of the sphere as the origin. This is extended to place solvent particles across the solute surface, point of contact between the sphere serving as the optimizing points. The angle between the two contact points, $\theta$, is determined as given in the fig.~\ref{fig-sph_pack}. The number of solvent molecule that can be tightly packed is obtained from the data \cite{sphcode}. The sphere onto which the points are optimally placed is an imaginary sphere which includes the critical radius as shown in the dashed lines. It is to be noted that this is still a theoretical estimate for contact packing on the imaginary larger sphere, and due to higher freedom of packing in our case, coordination states can be marginally higher for very large size ratios, $\sigma_s/\sigma_p \geq 5$. Table in fig.~\ref{fig-sph_pack} also gives the ma
 ximum an
 gle for which single bonding condition holds for a given size ratio ($\theta_{c,max}$) and specified critical distance ($r_c$).
 
 \begin{table}    
 	\begin{minipage}[b]{0.49\textwidth}
 		\centering
 		\begin{filecontents*}{Sphere_res.csv}
Ratio,rc,nmax,patchangle
0.5,0.80,8,17.64
0.8,0.99,12,23.83
0.9,1.05,12,25.51
1.0,1.10,14,27.04
1.1,1.16,16,28.44
1.2,1.21,18,29.73
2.0,1.55,31,40.18
\end{filecontents*}
\pgfplotstabletypeset[
 		col sep=comma,
 		header=has colnames,
 		string type,
 		columns/Ratio/.style = { column name = {$\sigma_s/\sigma_p$}},
 		columns/rc/.style = { column name = {$r_c$}},
 		columns/nmax/.style = { column name = {$n^{max}$}},
 		columns/patchangle/.style = { column name = {$\theta_{c,max}$},column type = c|},
 		every even column/.style = { column type = |c|},
 		every head row/.style={	before row= \hline, after row=\hline},
 		every last row/.style={after row=\hline}]{Sphere_res.csv}
 		\par\vspace{0pt}
 	\end{minipage}
 	\begin{minipage}[b]{0.49\textwidth}
 		\centering
 		\includegraphics[width=0.55\textwidth]{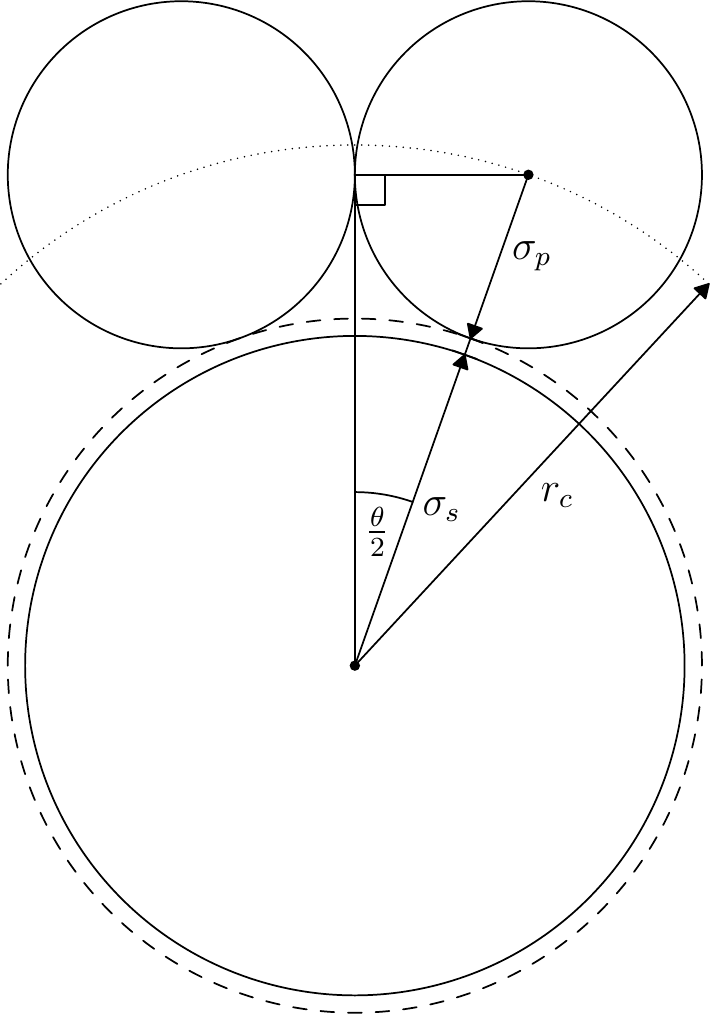} 
 		\par\vspace{0pt}
 	\end{minipage} 
 	\captionof{figure}{Left: Maximum number of coordination states possible and their respective maximum permissible patch angle for different size ratios $\sigma_s/\sigma_p$. Right: Diagram to compute the angle between two solvent particles to get the maximum number of coordination states}
 	 \label{fig-sph_pack}
 \end{table}
 
 Once the $n^{max}$ is defined, we find  $P^{( n )}$ as the probability that there is no hard sphere overlap  for randomly generated solvent molecules in the observation volume (or inner-shell) of solute molecules. As was discussed previously\cite{marshall_molecular_2013} that a hit-or-miss Monte Carlo \cite{hammersley,pratt_quasichemical_2001} approach to calculate $P^{(n)}$ proves inaccurate for large values of $n$ ($n>8$). But since 
 \begin{equation}
 {P^{( n)}} = P_{insert}^{( n )}{P^{( {n - 1} )}} \, , 
 \label{eq:12}
 \end{equation}
 where $P_{insert}^{(n)}$  is the probability of inserting a {\em{single}} particle given $n-1$ particles are already in the bonding volume, an 
 iterative procedure was be used to build the higher-order partition function from lower order one \cite{marshall_molecular_2013}.   The one-particle insertion probability $P_{insert}^{(n)}$ is easily evaluated using hit-or-miss Monte Carlo. Following is the table for isolated cluster probabilities for different size ratios.

 \begin{table}[!htbp]
 	\caption{Isolated cluster probabilities $P^{( n)}$ for different size ratios.}
 	\begin{scriptsize}
 		\centering
 		\begin{filecontents*}{isolated_cluster.csv}
n,0.8,0.85,0.9,0.95,1,1.05,1.1,1.15,1.2
1,1,1,1,1,1,1,1,1,1
2,0.72087,0.73574,0.74949,0.76219,0.77396,0.78485,0.79497,0.80444,0.81326
3,0.34732,0.37301,0.39762,0.42114,0.4435,0.46487,0.48523,0.50462,0.52305
4,0.10073,0.11924,0.13835,0.15792,0.1778,0.19765,0.21762,0.23745,0.25702
5,0.01518,0.02126,0.02854,0.03689,0.04632,0.05668,0.06793,0.07992,0.0926
6,9.61E-04,1.78E-03,3.03E-03,4.77E-03,7.10E-03,0.01008,0.01374,0.01812,0.02322
7,1.73E-05,5.28E-05,1.33E-04,2.89E-04,5.61E-04,9.97E-04,1.64E-03,2.56E-03,3.79E-03
8,6.04E-08,4.01E-07,1.86E-06,6.59E-06,1.91E-05,4.72E-05,1.03E-04,2.03E-04,3.70E-04
9,1.21E-10,4.25E-10,5.26E-09,3.98E-08,2.12E-07,8.64E-07,2.86E-06,7.95E-06,1.93E-05
10,1.17E-14,3.77E-13,3.56E-12,8.59E-11,1.48E-09,4.37E-09,2.74E-08,1.27E-07,4.67E-07
11,,1.73E-17,2.57E-15,5.50E-14,4.98E-12,3.87E-12,6.62E-11,6.46E-10,4.30E-09
12,,6.33E-24,2.60E-20,5.57E-18,1.11E-15,1.42E-15,4.78E-14,2.29E-12,1.27E-11
13,,,,,5.55E-24,4.44E-19,4.27E-17,1.07E-15,4.77E-15
\end{filecontents*}
\pgfplotstabletypeset[
 		col sep=comma,
 		header=has colnames,
 		columns/n/.style = { column name = {$n$} ,column type = |c|},
 		every even column/.style = { column type = c|},
 		every odd column/.style = { column type = c|},
 		every head row/.style={%
 			before row={					
 				\hline  &  \multicolumn{9}{c|}{$\sigma_s/\sigma_p$}  \\
 			},after row=\hline},
 		every last row/.style={after row=\hline}%
 		]{isolated_cluster.csv}
 	\end{scriptsize} 
 	\label{table:IC}       
 \end{table}

\newpage


\begin{thebibliography}{37}%
\makeatletter
\providecommand \@ifxundefined [1]{%
 \@ifx{#1\undefined}
}%
\providecommand \@ifnum [1]{%
 \ifnum #1\expandafter \@firstoftwo
 \else \expandafter \@secondoftwo
 \fi
}%
\providecommand \@ifx [1]{%
 \ifx #1\expandafter \@firstoftwo
 \else \expandafter \@secondoftwo
 \fi
}%
\providecommand \natexlab [1]{#1}%
\providecommand \enquote  [1]{``#1''}%
\providecommand \bibnamefont  [1]{#1}%
\providecommand \bibfnamefont [1]{#1}%
\providecommand \citenamefont [1]{#1}%
\providecommand \href@noop [0]{\@secondoftwo}%
\providecommand \href [0]{\begingroup \@sanitize@url \@href}%
\providecommand \@href[1]{\@@startlink{#1}\@@href}%
\providecommand \@@href[1]{\endgroup#1\@@endlink}%
\providecommand \@sanitize@url [0]{\catcode `\\12\catcode `\$12\catcode
  `\&12\catcode `\#12\catcode `\^12\catcode `\_12\catcode `\%12\relax}%
\providecommand \@@startlink[1]{}%
\providecommand \@@endlink[0]{}%
\providecommand \url  [0]{\begingroup\@sanitize@url \@url }%
\providecommand \@url [1]{\endgroup\@href {#1}{\urlprefix }}%
\providecommand \urlprefix  [0]{URL }%
\providecommand \Eprint [0]{\href }%
\providecommand \doibase [0]{http://dx.doi.org/}%
\providecommand \selectlanguage [0]{\@gobble}%
\providecommand \bibinfo  [0]{\@secondoftwo}%
\providecommand \bibfield  [0]{\@secondoftwo}%
\providecommand \translation [1]{[#1]}%
\providecommand \BibitemOpen [0]{}%
\providecommand \bibitemStop [0]{}%
\providecommand \bibitemNoStop [0]{.\EOS\space}%
\providecommand \EOS [0]{\spacefactor3000\relax}%
\providecommand \BibitemShut  [1]{\csname bibitem#1\endcsname}%
\let\auto@bib@innerbib\@empty
\bibitem [{\citenamefont
  {Wertheim}(1984{\natexlab{a}})}]{wertheim_fluids_1984}%
  \BibitemOpen
  \bibfield  {author} {\bibinfo {author} {\bibfnamefont {M.~S.}\ \bibnamefont
  {Wertheim}},\ }\href@noop {} {\bibfield  {journal} {\bibinfo  {journal} {J.
  Stat. Phys.}\ }\textbf {\bibinfo {volume} {35}},\ \bibinfo {pages} {19}
  (\bibinfo {year} {1984}{\natexlab{a}})}\BibitemShut {NoStop}%
\bibitem [{\citenamefont
  {Wertheim}(1984{\natexlab{b}})}]{wertheim_fluids_1984-1}%
  \BibitemOpen
  \bibfield  {author} {\bibinfo {author} {\bibfnamefont {M.~S.}\ \bibnamefont
  {Wertheim}},\ }\href@noop {} {\bibfield  {journal} {\bibinfo  {journal} {J.
  Stat. Phys.}\ }\textbf {\bibinfo {volume} {35}},\ \bibinfo {pages} {35}
  (\bibinfo {year} {1984}{\natexlab{b}})}\BibitemShut {NoStop}%
\bibitem [{\citenamefont {Chapman}\ \emph {et~al.}(1990)\citenamefont
  {Chapman}, \citenamefont {Gubbins}, \citenamefont {Jackson},\ and\
  \citenamefont {Radosz}}]{chapman_new_1990}%
  \BibitemOpen
  \bibfield  {author} {\bibinfo {author} {\bibfnamefont {W.~G.}\ \bibnamefont
  {Chapman}}, \bibinfo {author} {\bibfnamefont {K.~E.}\ \bibnamefont
  {Gubbins}}, \bibinfo {author} {\bibfnamefont {G.}~\bibnamefont {Jackson}}, \
  and\ \bibinfo {author} {\bibfnamefont {M.}~\bibnamefont {Radosz}},\
  }\href@noop {} {\bibfield  {journal} {\bibinfo  {journal} {Ind. Eng. Chem.
  Res.}\ }\textbf {\bibinfo {volume} {29}},\ \bibinfo {pages} {1709} (\bibinfo
  {year} {1990})}\BibitemShut {NoStop}%
\bibitem [{\citenamefont {Carnahan}\ and\ \citenamefont
  {Starling}(1969)}]{carnahan_equation_1969}%
  \BibitemOpen
  \bibfield  {author} {\bibinfo {author} {\bibfnamefont {N.~F.}\ \bibnamefont
  {Carnahan}}\ and\ \bibinfo {author} {\bibfnamefont {K.~E.}\ \bibnamefont
  {Starling}},\ }\href@noop {} {\bibfield  {journal} {\bibinfo  {journal} {J.
  Chem. Phys.}\ }\textbf {\bibinfo {volume} {51}},\ \bibinfo {pages} {635}
  (\bibinfo {year} {1969})}\BibitemShut {NoStop}%
\bibitem [{\citenamefont {Boubl{\'i}k}(1970)}]{boublik_hardsphere_1970}%
  \BibitemOpen
  \bibfield  {author} {\bibinfo {author} {\bibfnamefont {T.}~\bibnamefont
  {Boubl{\'i}k}},\ }\href@noop {} {\bibfield  {journal} {\bibinfo  {journal}
  {J. Chem. Phys.}\ }\textbf {\bibinfo {volume} {53}},\ \bibinfo {pages} {471}
  (\bibinfo {year} {1970})}\BibitemShut {NoStop}%
\bibitem [{\citenamefont {Mansoori}\ \emph {et~al.}(1971)\citenamefont
  {Mansoori}, \citenamefont {Carnahan}, \citenamefont {Starling},\ and\
  \citenamefont {{Leland~Jr.}}}]{mansoori_equilibrium_1971}%
  \BibitemOpen
  \bibfield  {author} {\bibinfo {author} {\bibfnamefont {G.~A.}\ \bibnamefont
  {Mansoori}}, \bibinfo {author} {\bibfnamefont {N.~F.}\ \bibnamefont
  {Carnahan}}, \bibinfo {author} {\bibfnamefont {K.~E.}\ \bibnamefont
  {Starling}}, \ and\ \bibinfo {author} {\bibfnamefont {T.~W.}\ \bibnamefont
  {{Leland~Jr.}}},\ }\href@noop {} {\bibfield  {journal} {\bibinfo  {journal}
  {J. Chem. Phys.}\ }\textbf {\bibinfo {volume} {54}},\ \bibinfo {pages} {1523}
  (\bibinfo {year} {1971})}\BibitemShut {NoStop}%
\bibitem [{\citenamefont {Bansal}\ \emph {et~al.}(2016)\citenamefont {Bansal},
  \citenamefont {Asthagiri}, \citenamefont {Cox},\ and\ \citenamefont
  {Chapman}}]{bansal_structure_2016}%
  \BibitemOpen
  \bibfield  {author} {\bibinfo {author} {\bibfnamefont {A.}~\bibnamefont
  {Bansal}}, \bibinfo {author} {\bibfnamefont {D.}~\bibnamefont {Asthagiri}},
  \bibinfo {author} {\bibfnamefont {K.~R.}\ \bibnamefont {Cox}}, \ and\
  \bibinfo {author} {\bibfnamefont {W.~G.}\ \bibnamefont {Chapman}},\
  }\href@noop {} {\bibfield  {journal} {\bibinfo  {journal} {J. Chem. Phys.}\
  }\textbf {\bibinfo {volume} {145}},\ \bibinfo {pages} {074904} (\bibinfo
  {year} {2016})}\BibitemShut {NoStop}%
\bibitem [{\citenamefont {Marshall}\ and\ \citenamefont
  {Chapman}(2013{\natexlab{a}})}]{marshall_molecular_2013}%
  \BibitemOpen
  \bibfield  {author} {\bibinfo {author} {\bibfnamefont {B.~D.}\ \bibnamefont
  {Marshall}}\ and\ \bibinfo {author} {\bibfnamefont {W.~G.}\ \bibnamefont
  {Chapman}},\ }\href@noop {} {\bibfield  {journal} {\bibinfo  {journal} {J.
  Chem. Phys.}\ }\textbf {\bibinfo {volume} {139}},\ \bibinfo {pages} {104904}
  (\bibinfo {year} {2013}{\natexlab{a}})}\BibitemShut {NoStop}%
\bibitem [{\citenamefont {Marshall}\ and\ \citenamefont
  {Chapman}(2013{\natexlab{b}})}]{marshall_thermodynamic_2013}%
  \BibitemOpen
  \bibfield  {author} {\bibinfo {author} {\bibfnamefont {B.~D.}\ \bibnamefont
  {Marshall}}\ and\ \bibinfo {author} {\bibfnamefont {W.~G.}\ \bibnamefont
  {Chapman}},\ }\href@noop {} {\bibfield  {journal} {\bibinfo  {journal} {Soft
  Matter}\ }\textbf {\bibinfo {volume} {9}},\ \bibinfo {pages} {11346}
  (\bibinfo {year} {2013}{\natexlab{b}})}\BibitemShut {NoStop}%
\bibitem [{\citenamefont {Torquato}(1986)}]{torquato_microstructure_1986}%
  \BibitemOpen
  \bibfield  {author} {\bibinfo {author} {\bibfnamefont {S.}~\bibnamefont
  {Torquato}},\ }\href@noop {} {\bibfield  {journal} {\bibinfo  {journal} {J.
  Stat. Phys.}\ }\textbf {\bibinfo {volume} {45}},\ \bibinfo {pages} {843}
  (\bibinfo {year} {1986})}\BibitemShut {NoStop}%
\bibitem [{\citenamefont {Reiss}\ \emph {et~al.}(1959)\citenamefont {Reiss},
  \citenamefont {Frisch},\ and\ \citenamefont
  {Lebowitz}}]{reiss_statistical_1959}%
  \BibitemOpen
  \bibfield  {author} {\bibinfo {author} {\bibfnamefont {H.}~\bibnamefont
  {Reiss}}, \bibinfo {author} {\bibfnamefont {H.~L.}\ \bibnamefont {Frisch}}, \
  and\ \bibinfo {author} {\bibfnamefont {J.~L.}\ \bibnamefont {Lebowitz}},\
  }\href@noop {} {\bibfield  {journal} {\bibinfo  {journal} {J. Chem. Phys.}\
  }\textbf {\bibinfo {volume} {31}},\ \bibinfo {pages} {369} (\bibinfo {year}
  {1959})}\BibitemShut {NoStop}%
\bibitem [{\citenamefont {Mayer}(1947)}]{mayer_integral_1947}%
  \BibitemOpen
  \bibfield  {author} {\bibinfo {author} {\bibfnamefont {J.~E.}\ \bibnamefont
  {Mayer}},\ }\href {\doibase 10.1063/1.1746468} {\bibfield  {journal}
  {\bibinfo  {journal} {J. Chem. Phys.}\ }\textbf {\bibinfo {volume} {15}},\
  \bibinfo {pages} {187} (\bibinfo {year} {1947})}\BibitemShut {NoStop}%
\bibitem [{\citenamefont {Torquato}\ and\ \citenamefont
  {Stell}(1982)}]{torquato_microstructure_1982}%
  \BibitemOpen
  \bibfield  {author} {\bibinfo {author} {\bibfnamefont {S.}~\bibnamefont
  {Torquato}}\ and\ \bibinfo {author} {\bibfnamefont {G.}~\bibnamefont
  {Stell}},\ }\href@noop {} {\bibfield  {journal} {\bibinfo  {journal} {J.
  Chem. Phys.}\ }\textbf {\bibinfo {volume} {77}},\ \bibinfo {pages} {2071}
  (\bibinfo {year} {1982})}\BibitemShut {NoStop}%
\bibitem [{\citenamefont {Torquato}\ and\ \citenamefont
  {Stell}(1983)}]{torquato_microstructure_1983}%
  \BibitemOpen
  \bibfield  {author} {\bibinfo {author} {\bibfnamefont {S.}~\bibnamefont
  {Torquato}}\ and\ \bibinfo {author} {\bibfnamefont {G.}~\bibnamefont
  {Stell}},\ }\href
  {http://scitation.aip.org/content/aip/journal/jcp/78/6/10.1063/1.445245}
  {\bibfield  {journal} {\bibinfo  {journal} {J. Chem. Phys.}\ }\textbf
  {\bibinfo {volume} {78}},\ \bibinfo {pages} {3262} (\bibinfo {year}
  {1983})}\BibitemShut {NoStop}%
\bibitem [{\citenamefont {Torquato}\ and\ \citenamefont
  {Stell}(1985)}]{torquato_microstructure_1985}%
  \BibitemOpen
  \bibfield  {author} {\bibinfo {author} {\bibfnamefont {S.}~\bibnamefont
  {Torquato}}\ and\ \bibinfo {author} {\bibfnamefont {G.}~\bibnamefont
  {Stell}},\ }\href
  {http://scitation.aip.org/content/aip/journal/jcp/82/2/10.1063/1.448475}
  {\bibfield  {journal} {\bibinfo  {journal} {J. Chem. Phys.}\ }\textbf
  {\bibinfo {volume} {82}},\ \bibinfo {pages} {980} (\bibinfo {year}
  {1985})}\BibitemShut {NoStop}%
\bibitem [{\citenamefont {Feng}\ and\ \citenamefont
  {Chapman}(2011)}]{feng_contact_2011}%
  \BibitemOpen
  \bibfield  {author} {\bibinfo {author} {\bibfnamefont {Z.}~\bibnamefont
  {Feng}}\ and\ \bibinfo {author} {\bibfnamefont {W.~G.}\ \bibnamefont
  {Chapman}},\ }\href@noop {} {\bibfield  {journal} {\bibinfo  {journal} {Mol.
  Phys.}\ }\textbf {\bibinfo {volume} {109}},\ \bibinfo {pages} {1717}
  (\bibinfo {year} {2011})}\BibitemShut {NoStop}%
\bibitem [{\citenamefont {Reiss}\ and\ \citenamefont
  {Merry}(1981)}]{reiss_upper_1981}%
  \BibitemOpen
  \bibfield  {author} {\bibinfo {author} {\bibfnamefont {H.}~\bibnamefont
  {Reiss}}\ and\ \bibinfo {author} {\bibfnamefont {G.~A.}\ \bibnamefont
  {Merry}},\ }\href@noop {} {\bibfield  {journal} {\bibinfo  {journal} {J.
  Phys. Chem.}\ }\textbf {\bibinfo {volume} {85}},\ \bibinfo {pages} {3313}
  (\bibinfo {year} {1981})}\BibitemShut {NoStop}%
\bibitem [{\citenamefont {Pratt}\ \emph {et~al.}(2001)\citenamefont {Pratt},
  \citenamefont {{LaViolette}}, \citenamefont {Gomez},\ and\ \citenamefont
  {Gentile}}]{pratt_quasichemical_2001}%
  \BibitemOpen
  \bibfield  {author} {\bibinfo {author} {\bibfnamefont {L.~R.}\ \bibnamefont
  {Pratt}}, \bibinfo {author} {\bibfnamefont {R.~A.}\ \bibnamefont
  {{LaViolette}}}, \bibinfo {author} {\bibfnamefont {M.~A.}\ \bibnamefont
  {Gomez}}, \ and\ \bibinfo {author} {\bibfnamefont {M.~E.}\ \bibnamefont
  {Gentile}},\ }\href@noop {} {\bibfield  {journal} {\bibinfo  {journal} {J.
  Phys. Chem. B.}\ }\textbf {\bibinfo {volume} {105}},\ \bibinfo {pages}
  {11662} (\bibinfo {year} {2001})}\BibitemShut {NoStop}%
\bibitem [{\citenamefont {Pratt}\ and\ \citenamefont
  {Ashbaugh}(2003)}]{pratt_selfconsistent_2003}%
  \BibitemOpen
  \bibfield  {author} {\bibinfo {author} {\bibfnamefont {L.~R.}\ \bibnamefont
  {Pratt}}\ and\ \bibinfo {author} {\bibfnamefont {H.~S.}\ \bibnamefont
  {Ashbaugh}},\ }\href@noop {} {\bibfield  {journal} {\bibinfo  {journal}
  {Phys. Rev. E}\ }\textbf {\bibinfo {volume} {68}},\ \bibinfo {pages} {021505}
  (\bibinfo {year} {2003})}\BibitemShut {NoStop}%
\bibitem [{\citenamefont
  {Wertheim}(1986{\natexlab{a}})}]{wertheim_fluids_1986}%
  \BibitemOpen
  \bibfield  {author} {\bibinfo {author} {\bibfnamefont {M.~S.}\ \bibnamefont
  {Wertheim}},\ }\href@noop {} {\bibfield  {journal} {\bibinfo  {journal} {J.
  Stat. Phys.}\ }\textbf {\bibinfo {volume} {42}},\ \bibinfo {pages} {459}
  (\bibinfo {year} {1986}{\natexlab{a}})}\BibitemShut {NoStop}%
\bibitem [{\citenamefont
  {Wertheim}(1986{\natexlab{b}})}]{wertheim_fluids_1986-1}%
  \BibitemOpen
  \bibfield  {author} {\bibinfo {author} {\bibfnamefont {M.~S.}\ \bibnamefont
  {Wertheim}},\ }\href@noop {} {\bibfield  {journal} {\bibinfo  {journal} {J.
  Stat. Phys.}\ }\textbf {\bibinfo {volume} {42}},\ \bibinfo {pages} {477}
  (\bibinfo {year} {1986}{\natexlab{b}})}\BibitemShut {NoStop}%
\bibitem [{\citenamefont {Beck}\ \emph {et~al.}(2006)\citenamefont {Beck},
  \citenamefont {Paulaitis},\ and\ \citenamefont {Pratt}}]{lrp:book}%
  \BibitemOpen
  \bibfield  {author} {\bibinfo {author} {\bibfnamefont {T.~L.}\ \bibnamefont
  {Beck}}, \bibinfo {author} {\bibfnamefont {M.~E.}\ \bibnamefont {Paulaitis}},
  \ and\ \bibinfo {author} {\bibfnamefont {L.~R.}\ \bibnamefont {Pratt}},\
  }\href@noop {} {\emph {\bibinfo {title} {The Potential Distribution Theorem
  And Models Of Molecular Solutions}}}\ (\bibinfo  {publisher} {Cambridge
  University Press},\ \bibinfo {address} {Cambridge, UK},\ \bibinfo {year}
  {2006})\BibitemShut {NoStop}%
\bibitem [{\citenamefont {Pratt}\ and\ \citenamefont
  {Asthagiri}(2007)}]{lrp:cpms}%
  \BibitemOpen
  \bibfield  {author} {\bibinfo {author} {\bibfnamefont {L.~R.}\ \bibnamefont
  {Pratt}}\ and\ \bibinfo {author} {\bibfnamefont {D.}~\bibnamefont
  {Asthagiri}},\ }in\ \href@noop {} {\emph {\bibinfo {booktitle} {Free Energy
  Calculations: {Theory} And Applications In Chemistry And Biology}}},\
  \bibinfo {series} {Springer series in {Chemical Physics}}, Vol.~\bibinfo
  {volume} {86},\ \bibinfo {editor} {edited by\ \bibinfo {editor}
  {\bibfnamefont {C.}~\bibnamefont {Chipot}}\ and\ \bibinfo {editor}
  {\bibfnamefont {A.}~\bibnamefont {Pohorille}}}\ (\bibinfo  {publisher}
  {Springer},\ \bibinfo {address} {Berlin, DE},\ \bibinfo {year} {2007})\
  Chap.~\bibinfo {chapter} {9}, pp.\ \bibinfo {pages} {323--351}\BibitemShut
  {NoStop}%
\bibitem [{\citenamefont {Merchant}\ and\ \citenamefont
  {Asthagiri}(2009)}]{merchant_thermodynamically_2009}%
  \BibitemOpen
  \bibfield  {author} {\bibinfo {author} {\bibfnamefont {S.}~\bibnamefont
  {Merchant}}\ and\ \bibinfo {author} {\bibfnamefont {D.}~\bibnamefont
  {Asthagiri}},\ }\href@noop {} {\bibfield  {journal} {\bibinfo  {journal} {J.
  Chem. Phys.}\ }\textbf {\bibinfo {volume} {130}},\ \bibinfo {pages} {195102}
  (\bibinfo {year} {2009})}\BibitemShut {NoStop}%
\bibitem [{\citenamefont {Merchant}\ \emph
  {et~al.}(2011{\natexlab{a}})\citenamefont {Merchant}, \citenamefont {Dixit},
  \citenamefont {Dean},\ and\ \citenamefont {Asthagiri}}]{merchant:jcp11b}%
  \BibitemOpen
  \bibfield  {author} {\bibinfo {author} {\bibfnamefont {S.}~\bibnamefont
  {Merchant}}, \bibinfo {author} {\bibfnamefont {P.~D.}\ \bibnamefont {Dixit}},
  \bibinfo {author} {\bibfnamefont {K.~R.}\ \bibnamefont {Dean}}, \ and\
  \bibinfo {author} {\bibfnamefont {D.}~\bibnamefont {Asthagiri}},\ }\href@noop
  {} {\bibfield  {journal} {\bibinfo  {journal} {J. Chem. Phys.}\ }\textbf
  {\bibinfo {volume} {135}},\ \bibinfo {pages} {054505} (\bibinfo {year}
  {2011}{\natexlab{a}})}\BibitemShut {NoStop}%
\bibitem [{\citenamefont {Merchant}\ \emph
  {et~al.}(2011{\natexlab{b}})\citenamefont {Merchant}, \citenamefont {Shah},\
  and\ \citenamefont {Asthagiri}}]{merchant_water_2011}%
  \BibitemOpen
  \bibfield  {author} {\bibinfo {author} {\bibfnamefont {S.}~\bibnamefont
  {Merchant}}, \bibinfo {author} {\bibfnamefont {J.~K.}\ \bibnamefont {Shah}},
  \ and\ \bibinfo {author} {\bibfnamefont {D.}~\bibnamefont {Asthagiri}},\
  }\href@noop {} {\bibfield  {journal} {\bibinfo  {journal} {J. Chem. Phys.}\
  }\textbf {\bibinfo {volume} {134}},\ \bibinfo {pages} {124514} (\bibinfo
  {year} {2011}{\natexlab{b}})}\BibitemShut {NoStop}%
\bibitem [{\citenamefont {Hummer}\ and\ \citenamefont
  {Szabo}(1996)}]{Hummer:jcp96}%
  \BibitemOpen
  \bibfield  {author} {\bibinfo {author} {\bibfnamefont {G.}~\bibnamefont
  {Hummer}}\ and\ \bibinfo {author} {\bibfnamefont {A.}~\bibnamefont {Szabo}},\
  }\href@noop {} {\bibfield  {journal} {\bibinfo  {journal} {J. Chem. Phys.}\
  }\textbf {\bibinfo {volume} {105}},\ \bibinfo {pages} {2004} (\bibinfo {year}
  {1996})}\BibitemShut {NoStop}%
\bibitem [{\citenamefont {Chang}\ and\ \citenamefont
  {Sandler}(1994)}]{chang_real_1994}%
  \BibitemOpen
  \bibfield  {author} {\bibinfo {author} {\bibfnamefont {J.}~\bibnamefont
  {Chang}}\ and\ \bibinfo {author} {\bibfnamefont {S.~I.}\ \bibnamefont
  {Sandler}},\ }\href@noop {} {\bibfield  {journal} {\bibinfo  {journal} {Mol.
  Phys.}\ }\textbf {\bibinfo {volume} {81}},\ \bibinfo {pages} {735} (\bibinfo
  {year} {1994})}\BibitemShut {NoStop}%
\bibitem [{\citenamefont {Yu}\ and\ \citenamefont
  {Karplus}(1988)}]{yu_thermodynamic_1988}%
  \BibitemOpen
  \bibfield  {author} {\bibinfo {author} {\bibfnamefont {H.-A.}\ \bibnamefont
  {Yu}}\ and\ \bibinfo {author} {\bibfnamefont {M.}~\bibnamefont {Karplus}},\
  }\href@noop {} {\bibfield  {journal} {\bibinfo  {journal} {J. Chem. Phys.}\
  }\textbf {\bibinfo {volume} {89}},\ \bibinfo {pages} {2366} (\bibinfo {year}
  {1988})}\BibitemShut {NoStop}%
\bibitem [{\citenamefont {Yu}\ \emph {et~al.}(1990)\citenamefont {Yu},
  \citenamefont {Roux},\ and\ \citenamefont {Karplus}}]{yu_solvation_1990}%
  \BibitemOpen
  \bibfield  {author} {\bibinfo {author} {\bibfnamefont {H.-A.}\ \bibnamefont
  {Yu}}, \bibinfo {author} {\bibfnamefont {B.}~\bibnamefont {Roux}}, \ and\
  \bibinfo {author} {\bibfnamefont {M.}~\bibnamefont {Karplus}},\ }\href@noop
  {} {\bibfield  {journal} {\bibinfo  {journal} {J. Chem. Phys.}\ }\textbf
  {\bibinfo {volume} {92}},\ \bibinfo {pages} {5020} (\bibinfo {year}
  {1990})}\BibitemShut {NoStop}%
\bibitem [{\citenamefont {Widom}(1982)}]{widom:jpc82}%
  \BibitemOpen
  \bibfield  {author} {\bibinfo {author} {\bibfnamefont {B.}~\bibnamefont
  {Widom}},\ }\href@noop {} {\bibfield  {journal} {\bibinfo  {journal} {J.
  Phys. Chem.}\ }\textbf {\bibinfo {volume} {86}},\ \bibinfo {pages} {869}
  (\bibinfo {year} {1982})}\BibitemShut {NoStop}%
\bibitem [{\citenamefont {Sivia}(1996)}]{sivia}%
  \BibitemOpen
  \bibfield  {author} {\bibinfo {author} {\bibfnamefont {D.}~\bibnamefont
  {Sivia}},\ }\href@noop {} {\emph {\bibinfo {title} {Data analysis -- A
  {Bayesian} tutorial}}}\ (\bibinfo  {publisher} {Oxford University Press},\
  \bibinfo {year} {1996})\BibitemShut {NoStop}%
\bibitem [{\citenamefont {Hummer}\ \emph {et~al.}(1998)\citenamefont {Hummer},
  \citenamefont {Garde}, \citenamefont {Garcia}, \citenamefont {Paulaitis},\
  and\ \citenamefont {Pratt}}]{lrp:jpcb98}%
  \BibitemOpen
  \bibfield  {author} {\bibinfo {author} {\bibfnamefont {G.}~\bibnamefont
  {Hummer}}, \bibinfo {author} {\bibfnamefont {S.}~\bibnamefont {Garde}},
  \bibinfo {author} {\bibfnamefont {A.~E.}\ \bibnamefont {Garcia}}, \bibinfo
  {author} {\bibfnamefont {M.~E.}\ \bibnamefont {Paulaitis}}, \ and\ \bibinfo
  {author} {\bibfnamefont {L.~R.}\ \bibnamefont {Pratt}},\ }\href@noop {}
  {\bibfield  {journal} {\bibinfo  {journal} {J. Phys. Chem. B.}\ }\textbf
  {\bibinfo {volume} {102}},\ \bibinfo {pages} {10469} (\bibinfo {year}
  {1998})}\BibitemShut {NoStop}%
\bibitem [{\citenamefont {Asthagiri}\ \emph {et~al.}(2003)\citenamefont
  {Asthagiri}, \citenamefont {Pratt},\ and\ \citenamefont
  {Kress}}]{asthagiri:pre03}%
  \BibitemOpen
  \bibfield  {author} {\bibinfo {author} {\bibfnamefont {D.}~\bibnamefont
  {Asthagiri}}, \bibinfo {author} {\bibfnamefont {L.~R.}\ \bibnamefont
  {Pratt}}, \ and\ \bibinfo {author} {\bibfnamefont {J.~D.}\ \bibnamefont
  {Kress}},\ }\href@noop {} {\bibfield  {journal} {\bibinfo  {journal} {Phys.
  Rev. E}\ }\textbf {\bibinfo {volume} {68}},\ \bibinfo {pages} {041505}
  (\bibinfo {year} {2003})}\BibitemShut {NoStop}%
\bibitem [{\citenamefont {Weisstein}(1999)}]{sphcode}%
  \BibitemOpen
  \bibfield  {author} {\bibinfo {author} {\bibfnamefont {E.~W.}\ \bibnamefont
  {Weisstein}},\ }\href@noop {} {\enquote {\bibinfo {title} {Spherical code},}\
  }\bibinfo {howpublished}
  {\url{http://mathworld.wolfram.com/SphericalCode.html}} (\bibinfo {year}
  {1999})\BibitemShut {NoStop}%
\bibitem [{\citenamefont {Sloane}\ \emph {et~al.}(1994)\citenamefont {Sloane},
  \citenamefont {Hardin},\ and\ \citenamefont {Smith}}]{sph_code2}%
  \BibitemOpen
  \bibfield  {author} {\bibinfo {author} {\bibfnamefont {N.~J.~A.}\
  \bibnamefont {Sloane}}, \bibinfo {author} {\bibfnamefont {R.~H.}\
  \bibnamefont {Hardin}}, \ and\ \bibinfo {author} {\bibfnamefont {W.~D.}\
  \bibnamefont {Smith}},\ }\href@noop {} {\enquote {\bibinfo {title} {Spherical
  codes},}\ }\bibinfo {howpublished} {\url{http://neilsloane.com/packings/}}
  (\bibinfo {year} {1994})\BibitemShut {NoStop}%
\bibitem [{\citenamefont {Hammersley}\ and\ \citenamefont
  {Handscomb}(1964)}]{hammersley}%
  \BibitemOpen
  \bibfield  {author} {\bibinfo {author} {\bibfnamefont {J.~M.}\ \bibnamefont
  {Hammersley}}\ and\ \bibinfo {author} {\bibfnamefont {D.~C.}\ \bibnamefont
  {Handscomb}},\ }\href@noop {} {\emph {\bibinfo {title} {Monte Carlo
  methods}}}\ (\bibinfo  {publisher} {Chapman and Hall},\ \bibinfo {address}
  {London},\ \bibinfo {year} {1964})\BibitemShut {NoStop}%
\end{thebibliography}

%

\end{document}